\documentclass[12pt,preprint]{aastex}

\def\HST{{\it HST}}

\def\Afsar{Af\c{s}ar}

\shorttitle{Geometric Distance for V838~Mon}
\shortauthors{Sparks et al.}

\begin{document}

\title{V838~Monocerotis: A Geometric Distance from {\it Hubble Space
Telescope} Polarimetric Imaging of its Light Echo\altaffilmark{1}}


\author{William B. Sparks,\altaffilmark{2} Howard E. Bond,\altaffilmark{2} Misty
Cracraft,\altaffilmark{2} 
Zolt Levay,\altaffilmark{2}		
Lisa A. Crause,\altaffilmark{3} 
Michael A. Dopita,\altaffilmark{4}	
Arne A. Henden,\altaffilmark{5}  	
Ulisse Munari,\altaffilmark{6}		
Nino Panagia,\altaffilmark{2,7}		
Sumner G. Starrfield,\altaffilmark{8}	
Ben E. Sugerman,\altaffilmark{9} 	
R. Mark Wagner,\altaffilmark{10} 
and 	
Richard L. White\altaffilmark{2}	
}

\altaffiltext{1}
{Based on observations with the NASA/ESA {\it Hubble Space Telescope\/} obtained
at the Space Telescope Science Institute, which is operated by the Association
of Universities for Research in Astronomy, Inc., under NASA contract
NAS5-26555.}

\altaffiltext{2}
{Space Telescope Science Institute, 3700 San Martin Drive, Baltimore, MD
21218; sparks@stsci.edu, bond@stsci.edu, cracraft@stsci.edu, levay@stsci.edu,
 rlw@stsci.edu}

\altaffiltext{3}
{South African Astronomical Observatory,
PO Box 9,
Observatory 7935,
Cape Town, South Africa}

\altaffiltext{4}
{RSAA, Australian National University,
Cotter Road, Weston Creek ACT 2611, Australia}

\altaffiltext{5}
{AAVSO, 49 Bay State Road, Cambridge, MA 02138 USA; arne@aavso.org}

\altaffiltext{6}
{INAF Osservatorio Astronomico di Padova, via dell'Osservatorio 8, 36012 Asiago,
Italy; ulisse.munari@oapd.inaf.it}

\altaffiltext{7}
{INAF/Osservatorio Astrofisico di Catania, Via S. Sofia 78, I-95123
Catania, Italy, and
Supernova Ltd., OYV \#131, Northsound Road, Virgin Gorda, British Virgin
Islands;
panagia@stsci.edu}

\altaffiltext{8}
{School of Earth and Space Exploration, Arizona State University, Tempe, AZ
85287; starrfield@asu.edu}

\altaffiltext{9}
{Dept.\ of Physics \& Astronomy, Goucher College,
1021 Dulaney Valley Rd.,
Baltimore, MD 21204;
ben.sugerman@goucher.edu
}

\altaffiltext{10}
{Large Binocular Telescope Observatory,
933 North Cherry Avenue,
Tucson, AZ 85721; rmw@as.arizona.edu}


\begin{abstract}

Following the outburst of the unusual variable star V838 Monocerotis in 2002, a
spectacular light echo appeared. A light echo provides the possibility of direct
geometric distance determination, because it should contain a ring of highly
linearly polarized light at a linear radius of $ct$, where $t$ is the time since
the outburst. We present imaging polarimetry of the V838~Mon light echo,
obtained in 2002 and 2005 with the Advanced Camera for Surveys onboard the {\it
Hubble Space Telescope}, which confirms the presence of the highly polarized
ring. Based on detailed modeling that takes into account the outburst light
curve, the paraboloidal echo geometry, and the physics of dust scattering and
polarization, we find a distance of $6.1\pm0.6$~kpc. The error is dominated by
the systematic uncertainty in the scattering angle of maximum linear
polarization, taken to be $\theta_{\rm max}=90^\circ \pm \, 5^\circ$. The
polarimetric distance agrees remarkably well with a distance of $6.2 \pm
1.5$~kpc obtained from the entirely independent method of main-sequence fitting
to a sparse star cluster associated with V838~Mon. At this distance, V838~Mon at
maximum light had $M_V\simeq-9.8$, making it temporarily one of the most
luminous stars in the Local Group. Our validation of the polarimetric method
offers promise for measurement of extragalactic distances using supernova light
echoes. 

\end{abstract}



\section{Introduction}

In early 2002, the previously unknown variable star V838~Monocerotis underwent
an unusual outburst. Its eruption light curve showed an initial brightening to
10th magnitude, followed about a month later by a dramatic rise to a sharp blue
peak at 6th magnitude. Over the ensuing nearly three months there were several
further episodes of re-brightening, after which it faded at optical wavelengths
to near its pre-outburst brightness. Unlike a classical nova, V838~Mon became
progressively redder, eventually becoming the coolest known luminous star (e.g.,
Bond et al.\ 2003; Evans et al.\ 2003; and papers in the proceedings of a
conference on the object, edited by Corradi \& Munari 2007).

Early in the outburst, a light echo surrounding V838~Mon was discovered by
Henden, Munari, \& Schwartz (2002). The echo phenomenon occurs when light from
an eruptive variable is scattered by dust in its vicinity, reaching the Earth at
progressively later times as the wave of illumination propagates outward.

Galactic light echoes are extremely rare.  The only other known example of
extent similar to that of V838~Mon was the echo produced by Nova GK Persei 1901
(Kapteyn 1902; Perrine 1902; Ritchey 1902). Following early misunderstandings,
light-echo geometry was properly described by Couderc (1939), and more recent
discussions are given by many authors, including Chevalier (1986), Felten
(1991), Sparks (1994), Sugerman (2003), and references therein. For an
instantaneous light flash, the echo surface at a given time is an ellipsoid with
the source at one focus and the Earth at the other. For astronomical distances,
this ellipsoid is well approximated by the paraboloid given by $z=x^2/2ct-ct/2$,
where $x$ is the projected distance from the star in the plane of the sky, $z$
is the distance from this plane along the line of sight (LOS) toward the Earth,
$c$ is the speed of light, and $t$ is the time since the outburst.  

The appearance of a light echo is governed by the time-dependent brightness of
the illuminating source, the density and scattering properties of the  dust, and
the distance to the star. Sparks (1994, 1996) showed that {\it polarimetric\/}
observations of a light echo may be used to derive a {\it purely geometric\/} 
distance to the source. While the intensity distribution of an echo can be
highly complex since it depends largely on the density of the scattering medium,
the linear-polarization distribution should be simple, smooth, and dependent
only on projected linear radius from the star and the time since the outburst.
This is true because the degree of linear polarization depends almost
exclusively on the scattering angle, which in turn is a monotonically decreasing
function of projected radius due to the parabolic geometry of a light echo. 

Furthermore, linear polarization due to scattering typically maximizes at a
scattering angle of $90^\circ$, i.e., for dust located in the plane of the sky.
As the above equation shows, the  $90^\circ$ scattering, occurring for dust at
$z=0$, takes place at a linear distance of $x=ct$ from the star. Hence there
should be a ring of linearly polarized light surrounding the source, whose
radius expands linearly with time, and whose angular radius yields the distance
geometrically. See Sparks (1994) for a detailed description of this novel method
for geometric determination of astronomical distances.

The spectacular light echo of V838~Mon thus offers a unique opportunity to test
this technique, and also to contribute to a physical understanding of this
extraordinary object by determining its distance and luminosity. 

\section{{\it Hubble Space Telescope\/} Observations}

We have obtained polarimetric images of the V838~Mon light echo using the Wide 
Field Channel (WFC) of the Advanced Camera for Surveys (ACS) onboard the {\it
Hubble  Space Telescope\/} ({\it HST\/})\null. Polarimetric observations were
made on five dates in 2002, and on one in 2005.  Table~1 presents an observing
log, including the dataset numbers in the {\it HST\/} archive.\footnote{The
\HST\/ data archive is available at http://archive.stsci.edu/hst} The
polarimetry was part of a larger, ongoing program of {\it HST\/} imaging of the
light echo, as described in Bond et al.\ (2003) and Bond (2007).

The polarimetric observations were made by imaging the echo through a bandpass
filter combined successively with three different polarizers. These polarizers
are oriented at relative position angles of $0^\circ$, $60^\circ$, and
$120^\circ$. The ACS has two sets of polarizers, one optimized for ultraviolet
wavelengths, and the other for visual. They are designated ``POL0UV,''
``POL60UV,'' and ``POL120UV'' for the ultraviolet, and ``POL0V,'' ``POL60V,''
and ``POL120V'' for the visual. The bandpass filter chosen for our first two
observations was the ACS ``$B$'' filter (F435W), and for the remaining four
observations it was the ``Broad $V$'' filter (F606W)\null. Each exposure through
each polarizer was repeated twice for cosmic-ray removal, and the exposure times
given in column~3 of Table~1 are the totals for the two exposures. For all of 
the $B$-band observations, the expected location of the linear-polarization ring
proved to be within a cavity in the dust around the star, and the $B$-band
images were therefore not analyzed further. The remainder of this paper deals
with the F606W $V$-band images, which were obtained at four different epochs.

The polarizing filters are relatively small, designed to fill only the field of
view (FOV) of the High-Resolution Channel (HRC) of the ACS\null. Therefore they
do not cover the entire FOV of the WFC\null. When the WFC is used, the
polarimetric images occupy approximately half the FOV of one of the two WFC
CCDs (i.e., one quadrant of a full WFC image), and only this quadrant is read
out at the end of each exposure.

In addition to the polarimetric images, we make use below of $B$- and $I$-band
direct images that were obtained contemporaneously with the  polarimetric
observations of 2002 December~17.  These direct images utilized the full WFC
FOV, and were used to create a three-color map of the echo for that epoch. On
2005 December~16 the light echo overfilled the FOV of the polarization quadrant
of the WFC, and so we also used the contemporaneous unpolarized full-field F606W
image obtained at the same epoch, in order to estimate the background sky level
underlying the polarimetric images, as described below. These unpolarized images
are listed in Table~2.

\section{Data Processing and Polarimetric Analysis}

\subsection{Pre-Processing Procedures}

The data were first processed through the standard Space Telescope Science
Institute (STScI) calibration pipeline in order to de-bias and flat-field the
individual images. In order simultaneously to remove cosmic rays, correct for 
geometric distortion, and register images spatially, we used the
STSDAS/PyRAF\footnote{The Space Telescope Science Data Analysis System (STSDAS)
and PyRAF are products of STScI, which is operated by AURA for NASA.}  task
``{\it multidrizzle}''\footnote{http://stsdas.stsci.edu/multidrizzle/} in
stand-alone mode. The final images were rectified onto a spatial grid  with
pixel size $0\farcs05\times0\farcs05$, and were registered to place V838~Mon at
the same location for each polarization sequence at each epoch. 

Because of a small optical magnification on the polarizers, the geometric
corrections are slightly different for each one. Small positional offsets are
also introduced by the polarizers, but then removed in the {\it HST\/}
pointing-control software; however, this procedure leads to small differences in
pipeline output image dimensions, and this too was adjusted in the stand-alone
{\it multidrizzle\/} runs so that the necessary image manipulations could be
performed on any given sequence of polarimetric observations. We retained a
detector coordinate system since the polarizer angles are fixed relative to the
instrument. {\it Multidrizzle\/} can be used with either exposure-time 
weighting (``EXP'') or photon-statistical weighting (``ERR'') (Pavlovsky et al.\
2006b). With just two images per pointing, the EXP weighting gives a robust,
unbiased estimate of the underlying count rate. The ERR weighting scheme biases
the flux estimate, since statistical  fluctuations in flux level propagate into
statistical fluctuations in the error estimate  (which is derived from the
flux). Empirical tests of both weighting schemes using  simulated noisy data
confirmed the bias in the ERR weighting, but demonstrated to our  satisfaction
the unbiased nature of the EXP weighting and validated our noise model described
below. Therefore, in the final data processing we used only the EXP weighting.

\subsection{Sky Background Subtraction}

Prior to the polarimetric analysis, we subtracted a constant background sky
level from each image. To estimate this underlying sky level, we measured the
mean level in twelve regions of each image well outside of, and distributed
around, the light echo. The local standard deviation of the sky was consistent
with our noise model; however there were small but significant differences
between the twelve measurements, corresponding to fluctuations in the background
of $\sim$3 to 5\%. Taking the mean of these measurements, we derived an
uncertainty in the sky level of approximately $\pm$0.001 counts~s$^{-1}$ for a
sky level that ranged from about 0.04 to 0.09~counts~s$^{-1}$.  Typical count
rates for faint portions of the light echo are generally around
0.1~counts~s$^{-1}$, so that uncertainties in the sky level should not be a
significant contributor to the error budget. For the 2005 December observation,
direct sky measurement was not possible because at that epoch the echo filled
the entire polarimetric FOV\null. However, at the same epoch we had also
obtained an unpolarized direct image in F606W using the entire FOV of the WFC,
as listed in Table~2. In this image sky measurement outside the echo is
straightforward. To estimate the sky level in the polarimetric images, we
registered the unpolarized image with the polarimetric image, and estimated the
sky backgrounds by plotting, on a pixel-by-pixel basis, the flux of the
unpolarized image against that of the polarized data. By using a linear fit, we
derived the sky level of each polarized image, knowing the sky level of the
unpolarized image.

\subsection{Polarimetric Analysis}

The output, sky-subtracted multidrizzled images were used to perform the
polarization analysis. As input to this analysis, we need the position angles of
the polarizers' electric vectors in detector coordinates, and the relative
throughputs of the three polarizers. For the polarizer position angles, we
adopted the ACS design specifications as given by Biretta et al.\ (2004, their
Table~19).  Note that Biretta \& Kozhurina-Platais (2004) have confirmed
empirically that the ACS position-angle design specifications were met closely
(within $0\fdg3$) in the flight instrument. For the relative throughputs we
adopted those given by Biretta et al.\ (2004, their Table~17) for the three
polarizers combined with the F606W filter. For convenience, we list the adopted
angles and throughputs in Table~3.  We then corrected the sky-subtracted count
rates for these relative throughputs.

Linear polarization is defined, at any point in an image, by the three Stokes
parameters, $I$, $Q$, and $U$, where $I$ is the total intensity, and $Q$ and $U$
are the linear-polarization components of the Stokes polarization vector. These
parameters can be derived using linear combinations of the three images obtained
with the polarizing elements (see, e.g., Collett 1992 and Tinbergen 1996 for
basic polarization concepts), in tandem with a noise model derived from the
data. We used the analysis procedure described by Sparks \& Axon (1999), which,
for perfectly efficient ideal polarizers oriented at relative position angles of
$60^\circ$, embodies the following conventions for the Stokes parameters:
$$I = (2/3) \, [ r({\rm POL0}) + r({\rm POL60}) + r({\rm POL120}) ] \, ,$$
$$Q = (2/3) \, [ 2r({\rm POL0}) - r({\rm POL60}) - r({\rm POL120}) ] \, ,$$
$$U = (2/\sqrt 3) \, [ r({\rm POL60}) - r({\rm POL120}) ] \, ,$$
where $r({\rm POL0})$, etc., are the source count rates in the three
polarimetric images.\footnote{The equation for $U$ is given as $U
= (2/\sqrt 3) \, [ r({\rm POL120}) - r({\rm POL60}) ]$ in  Biretta et al.\
(2004) and in the ACS Instrument Handbook (Pavlovsky et al.\ 2006a), but is
given correctly, as above, in the ACS Data Handbook (Pavlovsky et al.\ 2006b).}

From the Stokes parameters, we then derive the degree of linear polarization,
$p$, according to the convention
$$p = \sqrt{Q^2+U^2}/I\, ,$$ 
although, in detail, the Sparks \& Axon (1999) procedure also includes a
correction for the bias in these quantities introduced by their positive
definite nature. 

The polarization electric-vector position angle in detector coordinates,
$\theta_D$, is given by
$$\theta_D = (1/2) \tan^{-1}(U/Q) - 38\fdg2 \, ,$$
where the $-38\fdg2$ zero-point offset is the orientation of the POL0V
polarizer's electric vector projected onto the WFC detector, relative to the
spacecraft V3 axis, as listed in Table~3. To convert this electric-vector
position angle in the detector frame to an astronomical position angle on the
sky, $\theta_S$ (defined, as usual, with north at $0^\circ$ and east at
$90^\circ$), we used the formula
$$\theta_S = \theta_D + \theta_{\rm V3} \, ,$$
where $\theta_{\rm V3}$ is the position angle of the spacecraft V3 axis on the
sky, available in ACS images as the FITS image-header keyword ``PA\_V3.''

An important step in the polarization analysis is to establish a ``noise model''
(i.e., the relationship between intensity at each point in the image and its
uncertainty). This model is used to enable correction for the positive-definite
bias of polarization, due to the quadratic summation terms in the definition of
$p$.

To derive a noise model we started with the error images provided by the
pipeline at the ``flt'' or  ``crj'' stage (i.e., after flat-fielding but before
geometric correction). These error images are established using the standard
noise model described in Pavlovsky et al.\ (2004). We then drizzled the error
images to register them spatially with the drizzled data images, adjusting them
to units of count rates per second. At the geometrically corrected
(``drizzled'') stage, adjacent pixel values in images are spatially correlated
(Casertano et al.\ 2000). However, in order to  improve signal-to-noise, the
polarization analysis also bins  the data, which reduces this spatial
correlation from bin to bin. To estimate the smoothing introduced by these 
transformations taken together, we processed a simulated noise image containing
a standard normal distribution of pixel values. The amount by which the
simulated data were smoothed (determined empirically) was assumed to apply to 
the drizzled error image, and hence the estimated uncertainty on each pixel
intensity was determined. Polarization bias was then corrected for according to
the procedure described by Sparks \& Axon (1999), which yields a zero mean
polarization degree in the absence of source polarization. Uncertainties in the
derived polarization parameters are also provided in this processing.

As an overall verification of our procedures, we derived polarimetric properties
for a polarized star (Vela~I star~81; Bassino et al.\ 1982) observed in the
\HST\/ calibration program 10055 (P.I.: J.~Biretta). In this program, the star
was placed at the center of the ACS WFC FOV, and polarimetric observations were
taken at three different spacecraft roll angles ($\rm PA\_V3=348\fdg1, 48\fdg1,$
and $108\fdg0$) on three different dates in 2003-2004. Our analysis resulted in
polarization degrees of $p=5.8,$ 7.2, and 5.1\%, for a mean of 6.0\%. This can
be compared with the published ground-based measurement of $p=6.9\%$ (Whittet
et al.\ 1992). Our derived polarization position angles, $\theta_S$, were
$6\fdg4, 4\fdg3$, and $–0\fdg5$ for a mean of $\theta_S=3\fdg4$, vs.\ the
published $1\fdg0$. 

In a second phase of the same calibration program, on 2004 July 20-21, the
Vela~I star was placed at five different locations on the CCD, at a fixed roll
angle of $348\fdg1$.  The results at the five locations gave a very similar mean
of $p=5.7\%$ with a standard deviation of only 0.2\%, and a mean position
angle of $7\fdg5$ with standard deviation $0\fdg9$. These results suggest that
the variance seen between different roll angles is most likely due to
instrumental effects at the 1-2\% level in polarization degree and a few degrees
in position angle. 

A final check was performed in which we constructed a set of artificial ``flt''
images with associated Poisson noise, which spanned a range of surface
brightness, and contained a ring of  polarized light similar to that of the
V838~Mon light echo. Artificial ``error images'' were constructed  from the
noisy data, and the whole series was run through {\it multidrizzle\/} and our
polarization analysis software. For the exposure-time weighting scheme of {\it
multidrizzle\/}, the artificial noisy data properly reproduced both the
polarization pattern and its  estimated errors without any apparent bias.

\subsection{Polarimetry of the V838~Mon Images}

In Fig.~1 we show pictorial representations of the total-intensity (i.e., Stokes
$I$) images of V838~Mon at the four epochs of $V$-band \HST\/ polarimetric
imaging. The sequence of three images from 2002 shows the rapid apparent
increase in size of the echo from May through December, and by 2005 the echo was
larger than the FOV of the ACS polarimetry mode. For the 2002 images we placed
V838~Mon at the center of the FOV\null. In 2005, however, since we anticipated
that the diameter of the light-echo polarization ring would be larger than the
size of the FOV, we placed the star near a corner of the FOV.

In Fig.~2 we use image brightness to represent the linear-polarization degree,
$p$. Here we see that the complex structures seen in the intensity images of
Fig.~1 give way to a smooth, azimuthally symmetric morphology, just as expected
from a  straightforward scattering configuration. The 2002 May image shows that
the linear polarization increases inward toward the star, but does not reach a
maximum before encountering the prominent cavity surrounding the star. Hence
this image can be used only to derive an upper limit to the radius of the
polarization ring, and thus only a lower limit to the distance to the star. By
2002 September, it appears that the peak of the polarization ring has been
reached only on the south side of the star, but not elsewhere, where the cavity
radius measured from the central star is larger. By 2002 December, and also in
2005, there is dust present at many azimuths that contain the maximum of the
polarization ring. In the latter two images, the percent polarization at the
ring location is extremely high, approximately 50\%, as expected for light
scattered at $90^\circ$ off small particles (e.g., White 1979).

In Fig.~3 we represent the linear polarization in the image of 2002 December 17,
using red line segments, which are oriented to show the direction of the
electric vectors, and whose lengths are proportional to the degree of
polarization. Here we have left the image in detector coordinates and
orientation, rather than rotating it to place north at the top as in Figs.~1 and
2. The electric vectors are seen generally to be oriented perpendicular to the
direction to the central star. Again, this behavior is just as expected for dust
scattering, and it strongly confirms both the dust-scattering nature of the
light echo and our data-reduction procedures.\footnote{We thus generally do not
confirm the  distorted features seen in polarization maps presented by Desidera
et al.\ (2004, their Fig.~8), derived from ground-based polarimetric imaging of
V838~Mon on 2002 November~9.}

\subsection{Cleaning the Polarimetric Images}

For comparison to dust-scattering models of the light echo, and before using the
polarimetry to determine the distance to the star, it was necessary to carry out
two steps of cleaning the basic polarization images presented above. First, we
wished to accept only data of good signal-to-noise (S/N) ratio level. The
polarization uncertainties are provided by our noise model as described above,
and we chose to reject regions of the polarization-degree images having a
1$\sigma$ uncertainty in $p$ greater than 10\% of the value of $p$. In order to
retain an acceptable fraction of useful data, we first box-car smoothed the 2002
data using a $7\times7$-pixel box before the S/N-rejection process. In 2005
December the echo was much fainter, and we used $11\times11$-pixel smoothing.

Second, the presence of a substantial number of stars in the FOV complicates the
analysis. For each epoch, we generated a catalog of the field stars, measured
their fluxes, and generated an artificial star-field image from the catalog
using an empirical PSF derived from stars clear of the echo. Many of the stars
are significantly saturated, so our photometric procedure used a correction for
saturation based on the ratio of flux in an unsaturated annulus to the total
flux.  For the same reason, a composite empirical PSF was used that combined the
broad halo of a saturated star with the core region of a different star with
much reduced saturation. Then, following the same principles as in the previous
paragraph, we eliminated regions of the data where the stellar halos exceeded
10\% of the total Stokes $I$ intensity (which would introduce a 10\% error in
$p$). Beyond the stellar image cores, in the region where the halo has  less
than a 10\% influence on polarization, we subtracted the estimated stellar halo
from  the Stokes $I$ image and corrected the polarization-degree image
accordingly for this  small effect. Fig.~4 shows the result of masking out
regions with $\rm S/N<10$ and regions near bright stars, as described above, for
each of the four epochs.

\section{Modeling the Light Echo}

\subsection{Outburst Light Curve}

We next need to know the time behavior of the echo illumination due to the
outburst of V838~Mon. Fig.~5 shows the light curves that we adopted.  These are
ground-based Johnson-Kron-Cousins $B$, $V$, and $I$ magnitudes, which are
presented as a function of time. They are taken from a variety of ground-based
sources, and are the same data used in Bond et al.\ (2003), which can be
consulted for the original references. 

We then estimated the equivalent ACS count rates in the F435W, F606W, and F814W
filters, using the equations in Sirianni et al.\ (2005). These are the total
count rates from the star in electrons~s$^{-1}$ that ACS would have detected
throughout the eruption. 

To characterize the light curve and outburst, we derived several fiducial
characteristic quantities, as follows. For modeling purposes, we used only the
portion of the light curve between $(\rm HJD - 2400000)= 52307$ and 52404, that
is, the 97-day interval from 2002 February~2 to May~10. This interval excludes
the initial precursor rise prior to the sudden peak in early February; however
all of the flux emitted in the 32 days before February~2 constitutes only about
3.9\% of the total emitted flux, and its omission significantly reduces the
complexity of our analysis. Over the 97 outburst days, the derived average count
rate in the F606W plus POLV filter was $1.14\times 10^7$~counts~s$^{-1}$, and
the total counts emitted over the 97~days was $9.55\times10^{13}$. The 
flux-averaged mean Johnson $V$ magnitude is $\langle V\rangle=7.95$. The
peak-brightness Johnson $V$ magnitude is $V_{\rm max}=6.77$, with a peak ACS
F606W count rate of $2.89\times 10^7$ counts~s$^{-1}$. Thus the total  emitted
flux is equivalent to the $V_{\rm max}$ brightness shining for 38 days. The
intensity-weighted mean date of the outburst is HJD~2,452,343, or 2002 March~9.
Hence, the four epochs of our F606W polarimetric imaging correspond to times
after the outburst mean of about 72, 177, 282 and 1377 days. Note that, for the
first epoch, the outburst had ended, with the above conventions, only 9 days
earlier.

\subsection{Three-Dimensional Geometry and Appearance of the Light Echo}

As a prelude to the polarization analysis, it is useful to visualize the
geometry of the light echo at each of the epochs of observation.  Figure~6 shows
the geometric configuration at the four epochs. Anticipating the results below,
we show the locations of the light-echo paraboloids for an assumed distance of
6.1~kpc. The observer lies in the direction of the $+z$ axis.  The linear
distances on the $x$ axis have been converted to WFC pixels, whose $0\farcs05$
pixel scale corresponds to 0.0015~pc at the adopted distance. The $z$ distances
are shown in the same units.

Due to the protracted duration of the outburst, the dust illuminated at a given
epoch spans a significant depth in the LOS, especially at the earlier
epochs. For each of the four epochs, Figure~6 shows  light from the beginning of
the outburst as dashed lines, and light from the end (97 days later) as solid
lines. Along each LOS, the scattering  material is illuminated by a
replica of the complete light curve, which is somewhat distorted because of the
non-linear mapping of $t$ to $z$. 

Figure~6 shows that, for almost any assumed distribution of clumpy dust, the
LOS at a larger projected separation $x$ from the star corresponds to
light emitted earlier in the outburst. Thus, since the earliest light was much
bluer than light emitted toward the end of the outburst (see Figure~5), we
generally expect to see outer blue rims in the light echo.

Figure~7 shows a color rendition of the 2002 December~17 observations prepared
by one of us (Z. L.)  by combining the images in $B$ (F435W), $V$ (F606W), and
$I$ (F814W)\null.\footnote{The color representation is available at
http://hubblesite.org/newscenter/archive/releases/2005/02\slash image/f/} This
representation bears out beautifully the expectation described above, since it
shows numerous outer blue rims throughout the light echo. This color information
will allow us to determine three-dimensional depths for the dust clouds, which
will be done in a separate paper.

\subsection{Quantitative Source Functions for the Echo}

In order to model the polarization distribution of the echo, we began by
combining the light-curve data given in \S4.1 and plotted in Fig.~5 with the
three-dimensional geometry described in \S4.2 and illustrated in Fig.~6, to
produce a set of source functions. These are two-dimensional functions
(projected distance from the star and line-of-sight distance), with rotational
symmetry assumed for the third dimension. The source functions were then
integrated along the LOS to produce simulated polarization and intensity curves
for comparison with the actual data. 

For each LOS at each radius at each of the four epochs, we first established the
range of $z$ distances that corresponded to the beginning and end of the
outburst, as shown in Fig.~6. This $z$ range was then divided into equally
spaced sampling points, each of which corresponds to a particular time during
the outburst. Next we interpolated in the light curve to determine the flux
illuminating each of these $z$ volume elements (``voxels''). The illuminating
flux is given by $L(t)/4\pi d^2$, where $L(t)$ is the luminosity of the star at
time $t$ and $d$ is the distance of the voxel from the star. 

Figure~8 gives an illustration of the steps in the calculation of the source
functions. The top panel shows a representation of the illumination function for
the 2002 December 17 epoch. The horizontal axis is projected distance from the
star, and the vertical axis corresponds to distance along the $z$ axis toward
the observer, plotted on a linear distance scale from the largest time lag at
the bottom to the smallest lag at the top. As illustrated in Fig.~6, the
physical length along the $z$ axis increases from left to right. The thin,
nearly horizontal stripe at the bottom of the illumination function corresponds
to the initial sharp spike in the light curve, and then moving up we see the
subsequent dip in brightness, followed by a second brightening and then the
final fading. The distortion of the light curve, especially noticeable at small
projected $x$ distances, is due to the parabolic geometry of the echo.

Next we included the angular dependence of the efficiency of light scattering
off small particles. We adopted the usual Henyey-Greenstein (1941) phase
function 
$$\Phi(\theta) = (1-g^2)/[4\pi (1+g^2-2g\cos\theta)^{3/2}]\, ,$$
where $\theta$ is the scattering angle, with $\theta=0^\circ$ corresponding to
forward scattering. This function contains a parameter $g$, for which we adopt
the empirical value of $g=0.6$ (e.g., White 1979; Draine 2003). The second panel
in Figure~8 shows the effect of including the scattering phase function.  Here
we see the enhancement of the source function at larger values of $x$, where,
because of the parabolic geometry, we are closer to the highly efficient
forward-scattering regime. 

Finally, we included the dependence of degree of linear polarization on
scattering angle. We chose the classical polarization function for Rayleigh
scattering,
$$p=p_{\rm max} \, (1- \cos^2\theta)/(1+\cos^2\theta) $$ 
(e.g., Born \& Wolf 1980, section 13.5.2), which has its maximum value of
$p_{\rm max}$ at $\theta=90^\circ$. In the calculations below, we take $p_{\rm
max}$ to be a free parameter. For pure Rayleigh scattering, $p_{\rm max}=1$, but
in typical astronomical situations it is smaller. The bottom panel in Figure~8
shows the final source function with this factor included. We now see all of the
contributors along the LOS to the value of the Stokes $Q$ parameter.  The axial
symmetry of the echo geometry means there is no loss of generality from
computing only one of the two linear-polarization parameters, $Q$ and $U$\null. 

We can now integrate along the $z$ axis of each LOS, with the contributors to
the surface brightness in Stokes~$I$ given by
$$dI(z) = n_H(z)\,\omega\sigma  \{ L(t)\Phi(z)/(4\pi d^2 D^2) \}  dz \, ,$$ 
and those to Stokes $Q$ by
$$dQ(z) =  n_H(z)\,\omega \sigma p_{\rm max}
\{ L(t)\Phi(z)/(4\pi d^2 D^2) \}  
\{ (1-\cos^2\theta)/(1+\cos^2\theta) \} dz \, ,$$
where $n_H(z)$ is the volume density of hydrogen atoms, $\omega$ is the dust
albedo, and  $\sigma$ is the scattering cross-section per atom. We take
$\omega=0.6$ (Mathis, Rumpl, \& Nordsieck 1977) and $\sigma = 4.4\times
10^{-22}\,\rm cm^{2}$ at 6000~\AA\ (Savage \& Mathis 1979). $L(t)$
is the time-dependent luminosity of the star, and $\Phi(z)$ is the phase
function for the scattering angle $\theta$ corresponding to depth $z$. The terms
in curly braces may be recognized as the intensity, scattering, and polarization
source functions, with additional normalization for the distance $D$ of
V838~Mon. 

Quantitatively, we used units of ACS count rate for the term $L(t)/(4\pi D^2)$,
which is the flux that V838~Mon itself would have had, if it had been observed
throughout its outburst by the ACS\null. (This is the light curve computed in
\S4.1.) With this definition, summation of the voxels along each LOS yielded an
ACS count rate for each spatial pixel for an assumed constant density of $n_H =
1 \,\rm cm^{-3}$. Note that comparison of the integrated intensity with the
observed count rate yields an average density at each position, which will be
the subject of a separate paper.

To generate a polarization model for comparison with the data, we calculated the
contribution to Stokes~$Q$ for each voxel in a similar fashion, and also summed
these along the LOS, assuming constant density. Division of the integrated
Stokes~$Q$ function by the integrated intensity $I$ yielded a model polarization
degree, which was used for comparison with the data. Strictly speaking, the
polarization will depend on the density distribution, as this influences the
average intensity-weighted scattering angle along the LOS\null. However, these
integrations are only along a relatively narrow range in $z$, as shown in
Fig.~6. The total range of scattering angle along any given LOS for the 2002
December echo geometry is less than $20^\circ$, and is typically in the range
10--20$^\circ$.  Further, given a random distribution of dust clouds, we expect
the azimuthal averaging process to bring the actual average scattering angle
for a particular radius close to the idealized mean value of the models. A grid
of models was computed in this way for each epoch and for a range of distances
from 4 to 12~kpc.

\section{Geometric Distance}

\subsection{Polarization Profiles}

Figure~9 shows plots of the azimuthally averaged degree of polarization vs.\
radius for the F606W polarimetric observations obtained in 2002 September and
December. These curves were calculated from the fully cleaned $7\times7$-binned
polarization images, derived as described in \S\S3.3-3.5, and illustrated in
Figure~4. The azimuthal averages were calculated for annuli centered on
V838~Mon. 

The 2002 September data show a polarization maximum fairly close to the outer
edge of the central cavity around the star (see Fig.~1, upper-right panel). By
the time of the 2002 December observation, this maximum had moved to a larger
radius and had become well defined. The overall appearance of both the
polarization maps in Figure~4 and the azimuthally averaged polarization profiles
in Figure~9 is encouragingly smooth. This appearance is quite similar to
expectation (e.g., Sparks 1994, 2005), with a central dark hole, a smooth and
fairly wide maximum, and a shallow outer decline in polarization. The light echo
is about 50\% linearly polarized at the location of the maximum.

To derive a geometric distance, we begin with simple approximations and work
through to a detailed numerical analysis which takes into account the extended
duration of the outburst and the dust scattering and polarization properties.

\subsection{Approximate Estimates}

First, to obtain a sense of the approximate distance implied by these data, we
located the peaks of the polarization rings for 2002 September and December.
This was done by fitting parabolas to the peaks depicted in Figure~9, using only
polarization values greater than 0.4. The distance then follows by equating the
angular radius of the peak location to a linear radius of $ct$. We find the
respective peaks to lie at radii 99 and 167 pixels, corresponding to angular
radii of $4\farcs95$ and $8\farcs35$. Adopting an instantaneous approximation to
the outburst light curve, and delay times of $t=177$ and 282 days (cf.\ \S4.1),
these radii imply distances of $6.19\pm0.29$ and $5.85\pm0.13$~kpc, where the
uncertainties are the formal fitting errors and do not include systematic
uncertainties. The peak degrees of polarization are 48.5\% and 48.8\%,
respectively. 

The apparent outward motion of the polarization peak can also be used to obtain
a distance, without knowing the outburst date, by setting the angular motion to
a linear motion at the speed of light (e.g., Sparks 1997). The measured motion
of $3\farcs40$ in 105 days implies a distance of $5.36\pm0.47$~kpc. The
uncertainty is higher here since the errors on both individual measurements
combine quadratically.

\subsection{Detailed Modeling of the 2002 September and December Data}

Our detailed modeling concentrated initially on the observations of 2002
December.  We compared the full azimuthally averaged polarimetric data to model 
polarization curves determined numerically as described in \S4, for a range of
assumed distances $D$ and fitting $p_{\rm max}$ as a free parameter. Fig.~10
shows three examples of such fits to the 2002 December data. For each distance
we computed the rms residuals of the model fit to the polarization data, and
found a well-defined minimum corresponding to a distance $D=5.69$~kpc. The
best-fit value of $p_{\rm max}$ at this distance is 53.7\%.  This best model
polarization curve is the middle one plotted in Fig. 10, which also illustrates
the rapid degradation of fit quality away from the best-fit distance.  

Applying the same procedure to the 2002 September data, we found a very similar
best-fitting distance of 5.59~kpc, with $p_{\rm max} = 51.7\%$.  The best fits
to the 2002 September and December data are shown together in Figure~11.

To test the robustness of this approach, and reduce sensitivity to the shape of
the broad, low-polarization wings, we also derived rms residuals restricting the
models and data just to regions where the polarization degree was greater than
0.4, i.e., fitting the models just to the peak of the curve. For the 2002
December data, the best-fit distance increased slightly to 5.79~kpc from the
5.69~kpc of the fit to the whole curve, and for 2002 September it increased from
5.59 to 5.84 kpc. The peak polarizations are 51.3\% and 50.5\%, respectively.

One reason that the derived distance depends on the $p$ cutoff value is that, as
Fig.\ 11 shows, the observed polarization curves depart somewhat from the
Rayleigh-type model curves: the models have sharper and higher peaks, and they
decline more rapidly with radius than the actual data. Hence we turn to a
different way of looking at the data, which is to adopt a distance and then
derive the polarization degree as a function of scattering angle which is
implied by the data. In other words, we derive the empirical polarization phase
function that would be implied at each assumed distance. This approach makes use
of the fact that the assumed distance defines the relationship between {\it
angular radius\/} and {\it scattering angle}. Such phase curves have the
advantage that we expect them to be quite symmetric, unlike the significantly
asymmetric spatial polarization curves, given the exactly symmetric Rayleigh
phase function $p/p_{\rm max} = (1- \cos^2\theta)/(1+\cos^2\theta)$. A second
advantage is that we are now making a simpler and more direct comparison with a
basic physical property of the~dust.

For each numerical model, at each assumed distance, we calculated the
intensity-weighted mean scattering angle as a function of radius, taking into
account the geometry, light curve, and scattering functions described above. We
then plotted the observed degree of polarization against this mean scattering
angle. Fig.\ 12 shows examples of empirical scattering curves derived for assumed
distances of 4, 6, and 8 kpc from the 2002 December data. It is apparent from
Fig.\ 12 that the effect of assumed distance on the polarization scattering
function is indeed to move the location of the peak. For reference we also show
the Rayleigh phase function as a smooth curve in Fig.~12.

We fitted parabolas to the peaks of the phase functions, using only polarization
values $p>0.3$, in order to derive the locations of the peaks as a function of
assumed distance. The results are shown in Fig.~13, which again shows that the
derived distance decreases with increasing $\theta_{\rm max}$.  To estimate the
distance geometrically, we return to the basic assumption that polarization
maximizes at a scattering angle of $90^\circ$.  The 2002 December curve (solid
line in Fig.~13) intersects $90^\circ$ for a distance of 6.07~kpc.  The 2002
September data (dashed line) give a nearly identical distance of 6.12~kpc.
We may ask how robust is our assumption that the polarization maximum occurs at
$90^\circ$. Laboratory measurements of a variety of mineral dust particles
typically give polarization maxima in the range $90^\circ$--$100^\circ$ (Dumont
1973; Mu\~{n}oz, Volten, \& Hovenier 2002), while cometary dust is found
empirically to have a maximum polarization of about 30\% at about $90^\circ$
(Moreno, Mu\~{n}oz, \& Molina 2002 and references therein). Theoretically, White
(1979) used Mie theory to calculate the scattering properties of the Mathis et
al.\ (1977) grain distribution. His analysis yields $\theta_{\rm max}$ values
between $88^\circ$ and $95^\circ$ at wavelengths of 6000--6500~\AA\null. On this
basis we adopt $\theta_{\rm max}=90^\circ\pm5^\circ$ as a reasonable estimate of
the angle of maximum polarization and its systematic uncertainty.

In the case of an instantaneous outburst, it can be shown that the actual
distance is given exactly by $D = D_{90} (1+ \cos \theta_{\rm
max})/\sin\theta_{\rm max}$, where $\theta_{\rm max}$ is the actual scattering
angle for maximum polarization and $D_{90}$ is the distance derived assuming
$\theta_{\rm max}=90^\circ$. We overlay this function in Figure~13 for
$D_{90}=6.1$~kpc, where it may be seen that this analytical expression is
very close to the numerical model. In the neighborhood of $\theta_{\rm
max}=90^\circ$, the derived distance decreases by 1.7\% for each one-degree
increase in $\theta_{\rm max}$.

Thus, if the peak polarization were to occur $5^\circ$ on either side of
$90^\circ$, the derived distance would change by about 9\%, corresponding to a
distance range of 5.5 to 6.7~kpc.  This conclusion is robust to the cut-off
polarization degree assumed, as might be expected from the smooth and roughly
parabolic nature of the phase curves, as shown in Fig.~12. 

Next we tried the experiment of adopting the polarization phase function implied
by the 2002 September observations, and then using it to predict the spatial
distribution of polarization for the scattering-angle distribution at the time
of the 2002 December observation. The results are nearly independent of distance
and are shown in Fig.~14, which tests how consistent the polarization profiles
derived at the two epochs are. In fact, the consistency between the predicted
and observed curves is remarkably good. 

In Fig.~15 we show the polarization phase curves derived from the 2002 September
and December data for an adopted distance of 6.1~kpc, corresponding to
$\theta_{\rm max}=90^\circ$. The two phase curves are nearly identical, which
lends considerable confidence to our analysis methods.

As shown in Fig.~15, we fitted parabolas to the two phase functions, using
only the portions with $p>0.3$. We then used the parabolic phase functions to
predict the spatial polarization profiles at the 2002 September and December
epochs. Fig.~16 compares these predictions with the actual data. The fit to the
observed peaks is excellent. 

\subsection{The First and Fourth Epochs}

Polarization observations in the F606W filter were also obtained in 2002 May and
2005 December, as described in \S2 and summarized in Table~1. The 2002 May data
are unsuitable for distance measurement because the location of polarization
maximum was well within a cavity in the light echo at that epoch. 

The  data from 2005 December were also processed as described above. At that
late epoch, the echo was of much lower surface brightness than in 2002, and
relatively little of the image has polarization degree measured at high S/N, as
illustrated in Figs.~2 and 4. Hence a distance derived from such  data would be
liable to significant uncertainties. Instead, we content ourselves with a
comparison of the observed polarization to what is expected based on the results
of the late-2002 analysis. Fig.~17 shows the predicted polarization degree as a
function of projected radius. The predictions are based on the same
line-of-sight integrations described above, the Rayleigh formula for the
polarization phase function, and an assumed  distance of 6.1 kpc. We found that
a slightly higher maximum polarization degree of 0.55 instead of 0.5 appeared to
provide a better description of the data, but otherwise Fig.~17 shows that the
2005 data are reasonably consistent with the results obtained from the 2002
data. 

\section{The Distance and Nature of V838 Mon}

The uncertainty in our final distance estimate is dominated by systematic
errors, with the primarily source being the uncertainty in the scattering angle
that produces maximum linear polarization.  As discussed in \S5.3, we adopted
$\theta_{\rm max}=90^\circ\pm5^\circ$. This leads to our final best estimate of
$D=6.1\pm0.6$~kpc.\footnote{Sparks 2007, in a conference paper, presented a
preliminary analysis of our data, with a best estimate at that time of 5.9~kpc.
This value is now superseded since, as described in the present work, we have
improved the {\it multidrizzle\/} weighting scheme used to prepare the
polarimetric images, done a better job of estimating and removing field-star
contamination, improved the sky subtraction, and completed our analysis of the
polarization phase curves from all epochs.}

Initial estimates of the distance to V838~Mon ranged over more than an order of
magnitude. Shortly after the appearance of the light echo, a very short distance
($\sim$0.7~kpc) was derived under the assumption that the {\it outer\/} edge of
the echo lies at a projected radius of $ct$ (e.g., Munari et al.\ 2002;
Kimeswenger et al.\ 2002). As explained by Munari et al.\ (2005), the implicit
assumption had been made that the dust formed a face-on disk around the star,
but we now know that this is not the case. In fact, the apparent expansion rate
of a light echo is almost always superluminal. Using the proper paraboloidal
geometry, Bond et al.\ (2003) showed, from the apparent expansion rates of the
echo, that the distance must be substantially larger than 2~kpc.

As V838~Mon declined in visual light in late 2002, spectroscopic observations
revealed an unresolved B3~V companion star (Munari \& Desidera 2002; Wagner \&
Starrfield 2002). Giving high weight to a spectroscopic parallax of the B3
companion, Munari et al.\ (2005) favored a much larger distance of $\sim$10~kpc.
However, \Afsar\ \& Bond (2007) have argued that this is an overestimate because
the B star suffers extinction over and above that due to the foreground
interstellar medium; this extra extinction would be due to the B~star lying
within (or behind) dust ejected from V838~Mon during the 2002 outburst. This
argument appears to have been confirmed by recent dramatic episodes of fading of
light from the B3 star (Goranskij 2006; Bond 2006; Munari et al.\ 2007), attributed to dust from
V838~Mon engulfing or passing in front of it.

A more reliable distance comes from the discovery by \Afsar\ \& Bond (2007) that
V838 Mon belongs to a sparse young cluster, containing three B-type stars in
addition to the unresolved B3 companion. By photometric main-sequence fitting of
the three B stars, \Afsar\ \& Bond find a reddening of $E(B-V)=0.85$ and a
distance of $6.2\pm1.2$~kpc. The agreement with the polarimetric distance of
$6.1 \pm 0.6$~kpc is excellent. We are attempting to identify more members of
the cluster in order to refine the distance estimate, which would provide
additional support for our assumption of $\theta_{\rm max}=90^\circ$ in the
polarization analysis.

The apparent magnitude of V838~Mon at its maximum in early 2002 February was
$V=6.77$ (see Fig.~5). At a distance of 6.1~kpc, and for the reddening given
above, the absolute magnitude at maximum was an extraordinary $M_V=-9.8$, making
V838~Mon temporarily one of the visually brightest stars in the entire Local
Group (e.g., Humphreys 1983, Table~3). V838~Mon was brighter at maximum than all
but the very fastest of classical novae (e.g., Downes \& Duerbeck 2000, their
Fig.~17).

In 1988, a luminous cool object appeared in the nuclear bulge of M31. This
object, called the ``M31 red variable'' or ``M31 RV,'' is widely considered to
be an analog of V838~Mon (e.g., Bond \& Siegel 2006 and references therein). 
Photometric coverage of its outburst, collected by Boschi \& Munari (2004, their
Table~5), was unfortunately fairly sparse, especially in the $V$ band. However, 
by combining the brightest observed $B$ magnitude, $B=17.3$, with a color index
of $B-V=1.89$ observed later in the outburst, we can estimate $V_{\rm
max}\simeq15.4$. Adopting $E(B-V)=0.12$ and $(m-M)_0=24.48$ for M31~RV from
Boschi \& Munari, we find $M_{V\rm,max}\simeq-9.4$, similar to the maximum
brightness of V838~Mon. 

The Galactic star V4332~Sgr is a third object that was also a red supergiant
throughout its outburst (Martini et al.\ 1999; Tylenda et al.\ 2005 and
references therein). Its distance is, however, poorly constrained.

The outburst mechanism for this new class of objects remains uncertain. Their
outburst behavior, featuring a rapid expansion to a very cool, luminous
supergiant, is unlike that of any previously known type of variable star. A
thermonuclear runaway on a white dwarf in a nova-like system is probably ruled
out by the very young age of the cluster surrounding V838~Mon, $<$25~Myr
(\Afsar\ \& Bond 2007), which appears to be too short to allow formation of a
close binary containing a white dwarf. An explosive event in a massive star
seems to be excluded by the fact that the population surrounding M31~RV contains
only old red giants (Bond \& Siegel 2006). The high luminosities and rapid
expansion timescales appear to disallow a ``born-again'' event in a low-mass
pre-white dwarf.

Thus recent discussions of these objects have concentrated on stellar-merger
scenarios (e.g., Tylenda \& Soker 2006 and references therein) and even
planet-star mergers (e.g., Retter et al.\ 2006). Mergers have the advantage of
possibly occurring in both young and old populations. Moreover, they lead to an
expectation of a wide range of outburst luminosities, determined by the
potentially wide range of parameters of the merging objects. This expectation
appears to be borne out by the recent discovery of an apparently V838~Mon-like
event in the Virgo galaxy M85 (Rau et al.\ 2007; Kulkarni et al.\ 2007),
which, at a peak absolute magnitude of $M_V\simeq-13$, was at least $\sim$20
times more luminous than V838~Mon and M31~RV at maximum.

A related issue, which also bears on outburst mechanisms for V838~Mon, is the
origin of the dust being illuminated in the light echo. Different authors have
argued that the dust was ejected from V838~Mon in a previous outburst, or that
it is ambient interstellar dust whose origin is unrelated to V838~Mon itself. If
the dust did come from a previous outburst, it would cast doubt on the
stellar-merger scenarios, which presumably would be one-time events.  These
issues do not affect our distance determination, and are thus beyond the scope
of the present paper; see the recent conference proceedings for discussions of
the origin of the dust (Corradi \& Munari, eds., 2007).

\section{The Utility of Light Echoes for Geometric Distance Measurement}

The use of light-echo polarimetry to obtain geometric distances of supernovae
(SNe) was proposed by Sparks (1994). The method has now found its application to
the distance to an extraordinary outburst event within the Milky Way. The result
agrees extremely well with the distance obtained from a completely independent,
classical method of cluster main-sequence fitting, thus providing strong support
for the validity of the polarimetric technique.

SN light echoes offer the prospect of geometric distance measurements on an
extragalactic scale. In fact, SNe form a more favorable class of object than
V838~Mon, because of the short outburst timescale, and because SNe are
intrinsically extremely bright and can potentially illuminate bright,
long-duration light echoes. However, there must be interstellar dust distributed
sufficiently in the vicinity of the SN to produce both a light echo and a clear
maximum of linear polarization, corresponding to the dust lying in the plane of
the sky.

The angular diameter of the linear-polarization ring in a SN light echo
(corresponding to a linear diameter of $2ct$) is given by $\theta = 0\farcs126
\, (t/{\rm 1 \,yr}) \, ({\rm 1 \, Mpc}/D)$. Such sizes make the polarimetric
method feasible with ground-based techniques within and perhaps slightly outside
the Local Group. At {\it HST\/} resolution, geometric distance measurements to
galaxies within several Mpc should be feasible.\footnote{At this writing the
Wide Field Channel of ACS onboard {\it HST\/} is inoperable, but it is possible
that ACS operations will be restored in the servicing mission currently planned
for 2008.}  In fact, linear polarization was detected in {\it HST\/} polarimetry
of a light echo around SN~1991T in the Virgo galaxy NGC~4527; modeling of the
polarization (Sparks et al.\ 1999) showed consistency with a distance of
$\sim$15~Mpc, but these images, taken at $t\simeq6$--8~yr after the outburst,
only marginally resolved the echo.

\section{Conclusions}

We have applied the method of light-echo polarimetric imaging described by
Sparks (1994) to the derivation of a geometric distance to V838~Mon. Following a
careful reduction of polarimetric images of the light echo, obtained with the
Advanced Camera for Surveys onboard the {\it Hubble Space Telescope}, we
confirmed the presence of an apparently expanding ring of highly polarized light
surrounding V838~Mon. Setting the radius of this ring to a linear size of $ct$
yielded initial distance estimates of 5.4--6.2~kpc. We then developed modeling
and fitting procedures of increasing complexity, which yielded consistent
results. 

Our final best value for the distance is $6.1\pm0.6$~kpc, where the error is
dominated by the systematic uncertainty in the scattering angle of maximum
linear polarization. This distance agrees extremely well with a value of
$6.2\pm1.2$~kpc, obtained by an entirely independent method of main-sequence
fitting, applied to a sparse stellar cluster to which V838~Mon belongs.  

At this distance, the outburst of V838~Mon in early 2002 is shown to have been
extremely luminous---it was brighter than all but the most luminous classical
novae, and temporarily one of the brightest stars in the entire Local Group.
The mechanism that produced such an outburst remains uncertain, but scenarios
involving stellar mergers may be the most plausible.

Our results strongly verify the polarimetric method for determining geometric
distances. The technique shows great promise for application to distance
determinations for extragalactic supernovae.

\acknowledgments

Support for this project was provided by NASA through grants from the Space
Telescope Science Institute, which is operated by the Association of
Universities for Research in Astronomy, Inc., under NASA contract NAS5-26555.
S.S. acknowledges partial support from NSF and NASA grants to Arizona State
University.

\clearpage

\begin{figure}
\begin{center}
\includegraphics[width=6.5in]{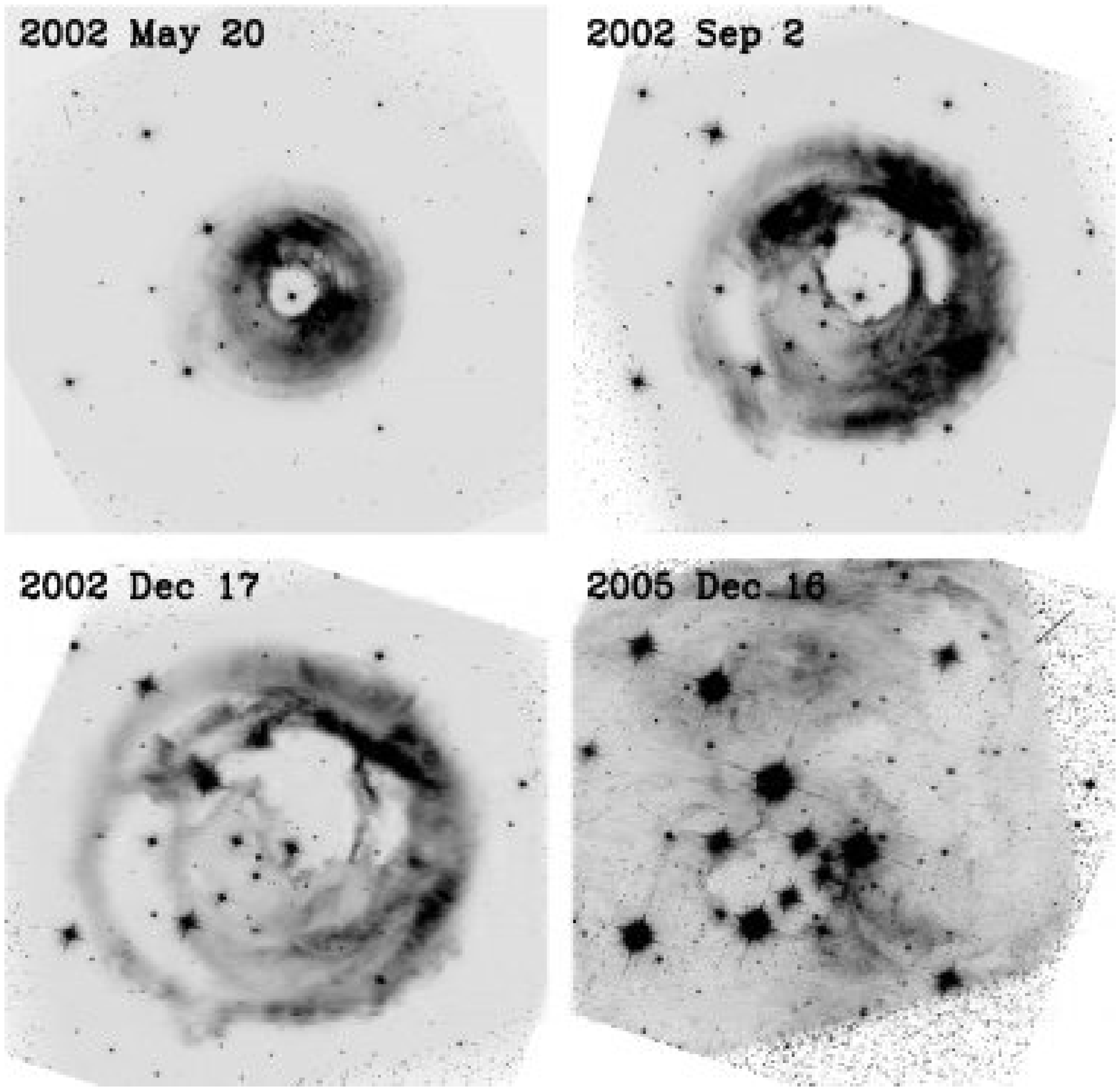}
\end{center}
\figcaption{
Total-intensity (i.e., Stokes $I$) \HST\/ ACS images of the V838~Mon light echo
in the $V$ (F606W) filter for each of the four epochs of polarimetric data. The
images have been rotated to place north up and east on the left, and each frame
is $97''$ wide. Polarimetric observations were obtained on 2002 May 20,
September 2, and December 17, and on 2005 December 16, as labelled in the
figure.  The image stretch is linear, and the fading of the light echo has been
compensated by scaling the image intensities. V838~Mon itself is located near
the center of the cavity in the 2002 May 20 image.
}
\end{figure}

\begin{figure}
\begin{center}
\includegraphics[width=6.5in]{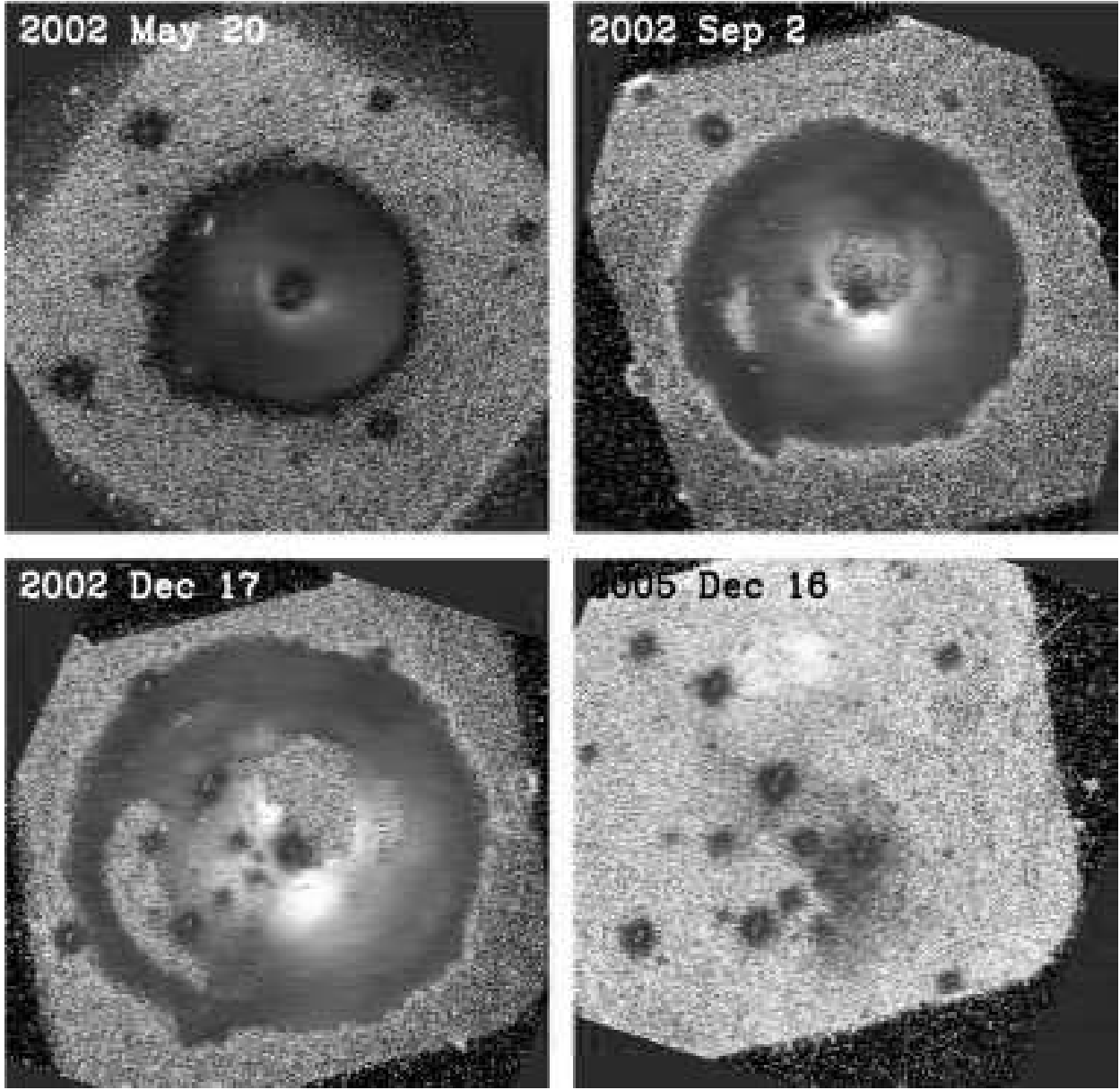}
\end{center}
\figcaption{
Images representing the degree of linear polarization, $p$, for each of the
four epochs of data shown in Fig.~1. Image scales and orientations are the same
as in Fig.~1. The image stretch is linear, ranging from black representing zero
linear polarization to full white representing $\sim$50\% linear polarization.
These images illustrate the apparent outward motion of a ring of highly
polarized light in the light echo. 
}
\end{figure}

\begin{figure}
\begin{center}
\includegraphics[width=6in]{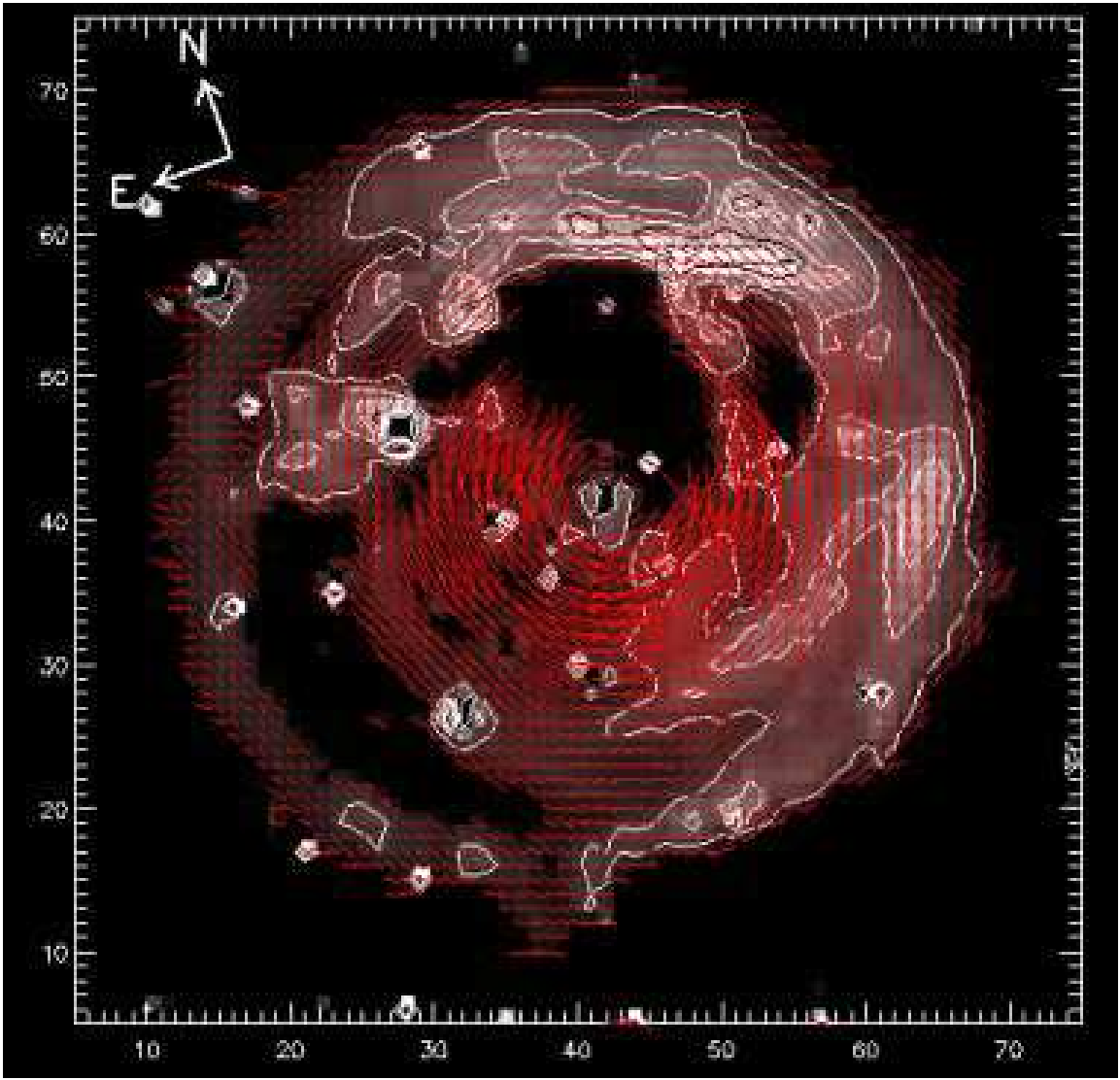}
\end{center}
\figcaption{
The echo image of 2002 December 17 (contours and greyscale image) with
polarization electric vectors  superimposed. The directions of the electric
vectors are indicated by the red lines, whose lengths are proportional to the
degree of polarization. The largest values are about 50\%. Vectors are shown
every 30 pixels, and the polarization and position angles are the means and
medians, respectively, averaged over $30\times30$-pixel boxes. This image has
{\it not\/} been rotated to place north at the top, but instead remains in
detector coordinates. Small tickmarks on the axes are separated by $1\farcs5$.
Note that the electric vectors are generally perpendicular to the direction to
the central star, as expected for light scattered off dust particles. 
} \end{figure}

\begin{figure}
\begin{center}
\includegraphics[width=6.5in]{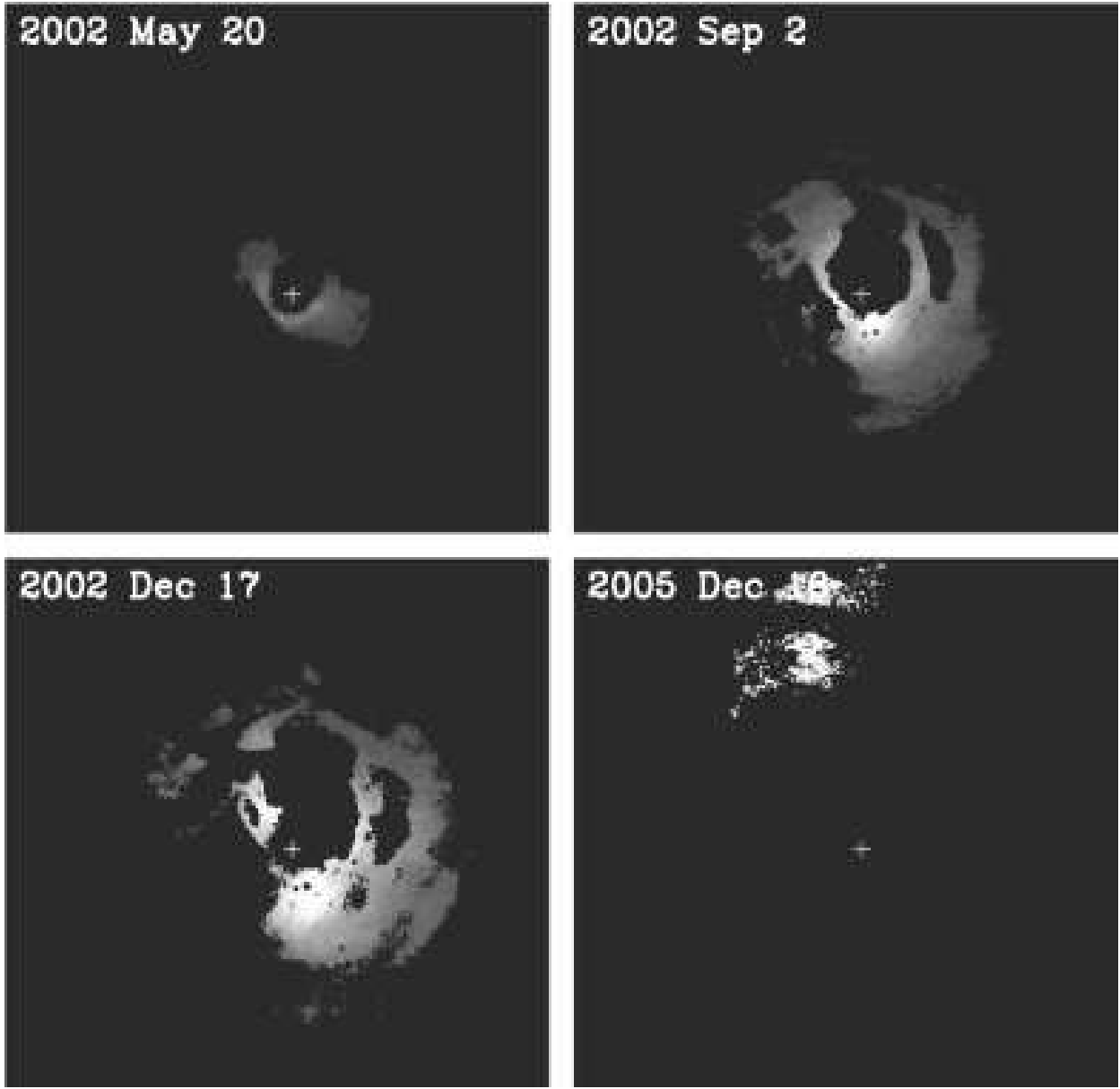}
\end{center}
\figcaption{
The images illustrating polarization degree shown in Fig.~2 are here masked to
retain only regions free of stars and with linear polarization degree detected
with $\rm S/N>10$, as described in the text. Image scale, orientation, and
stretch are the same as in Fig.~2. The location of V838~Mon in each masked
image is marked with a white cross. 
}
\end{figure}

\begin{figure}
\begin{center}
\includegraphics[width=6in]{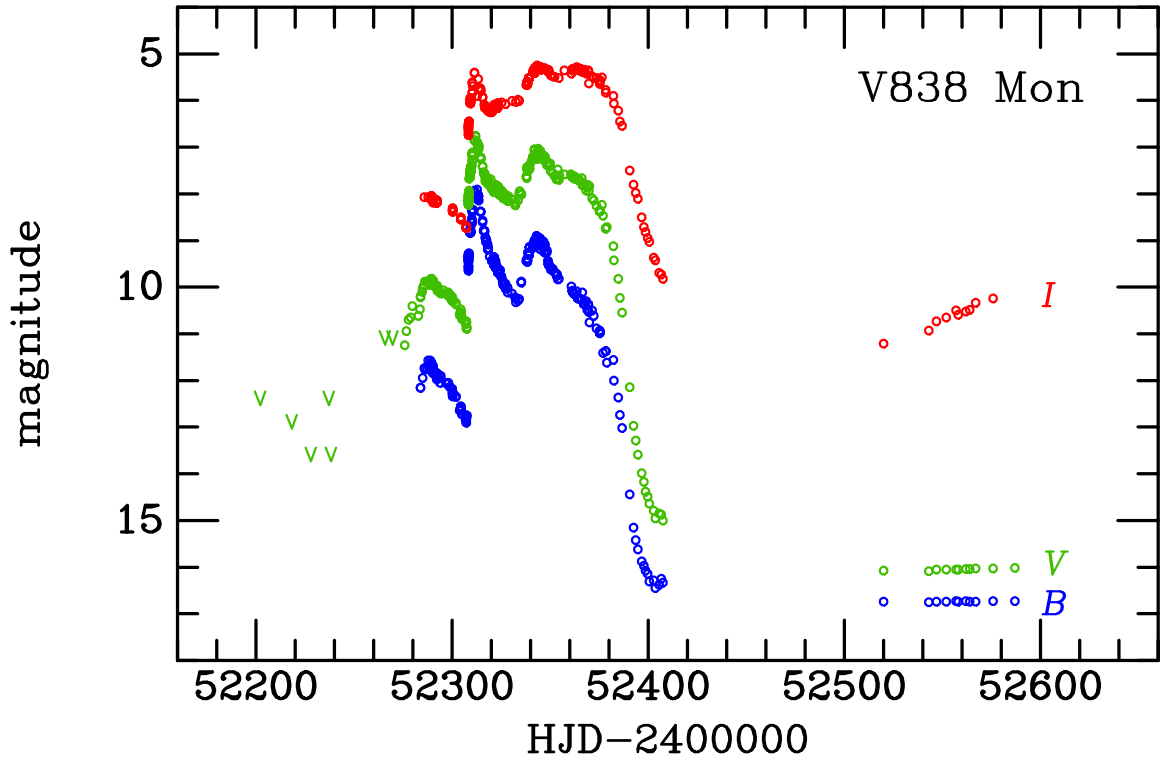}
\end{center}
\figcaption{
Outburst light curves adopted for V838~Mon in the ground-based
Johnson-Kron-Cousins $B$, $V$, and $I$ bandpasses, colored blue, green, and red
respectively. ``V'' signs mark upper limits before the onset of the outburst.
The data are the same as used by Bond et~al.\ (2003), but are plotted here as
stellar magnitudes rather than on a linear flux scale. 
}
\end{figure}

\begin{figure}
\begin{center}
\includegraphics[width=6in]{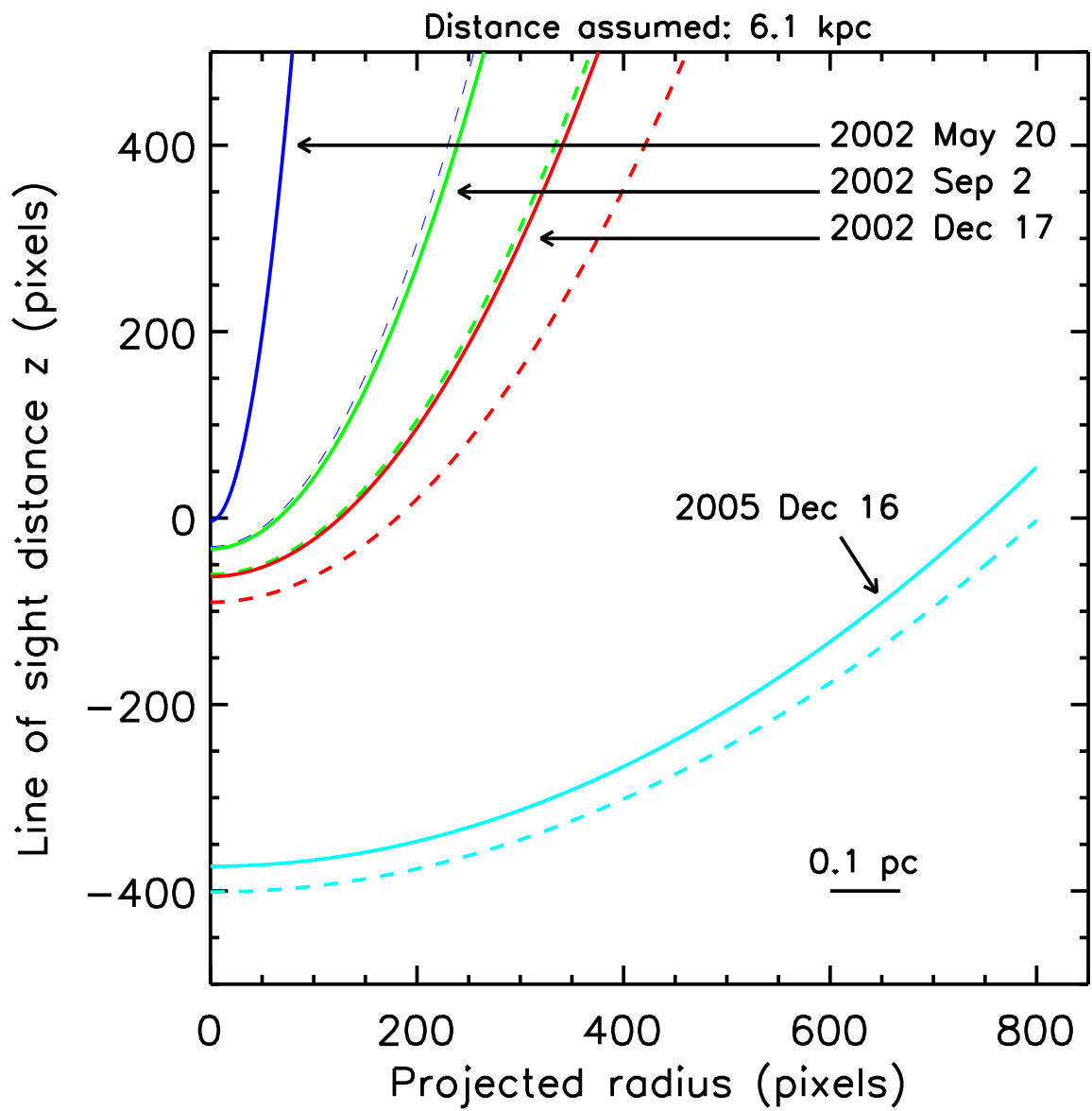}
\end{center}
\figcaption{
Spatial locations of the light-echo paraboloids for our four epochs of
F606W polarimetric imaging. The observer is located on the $+z$ axis,
V838~Mon lies at the (0,0) origin, and the $x$ and $z$ distances from the star
are in units of ACS pixels ($0\farcs05$) for an adopted distance of 6.1~kpc. The
850-pixel width of the diagram corresponds to a linear width of 1.26~pc. At each
epoch, the solid line marks light from the end of the outburst (shortest time
delay), and the dashed line marks the beginning of the outburst (longest time
delay). Blue represents the paraboloids for 2002 May, green for 2002 September,
red for 2002 December, and cyan for 2005 December.
}
\end{figure}

\begin{figure}
\begin{center}
\includegraphics[width=6.5in]{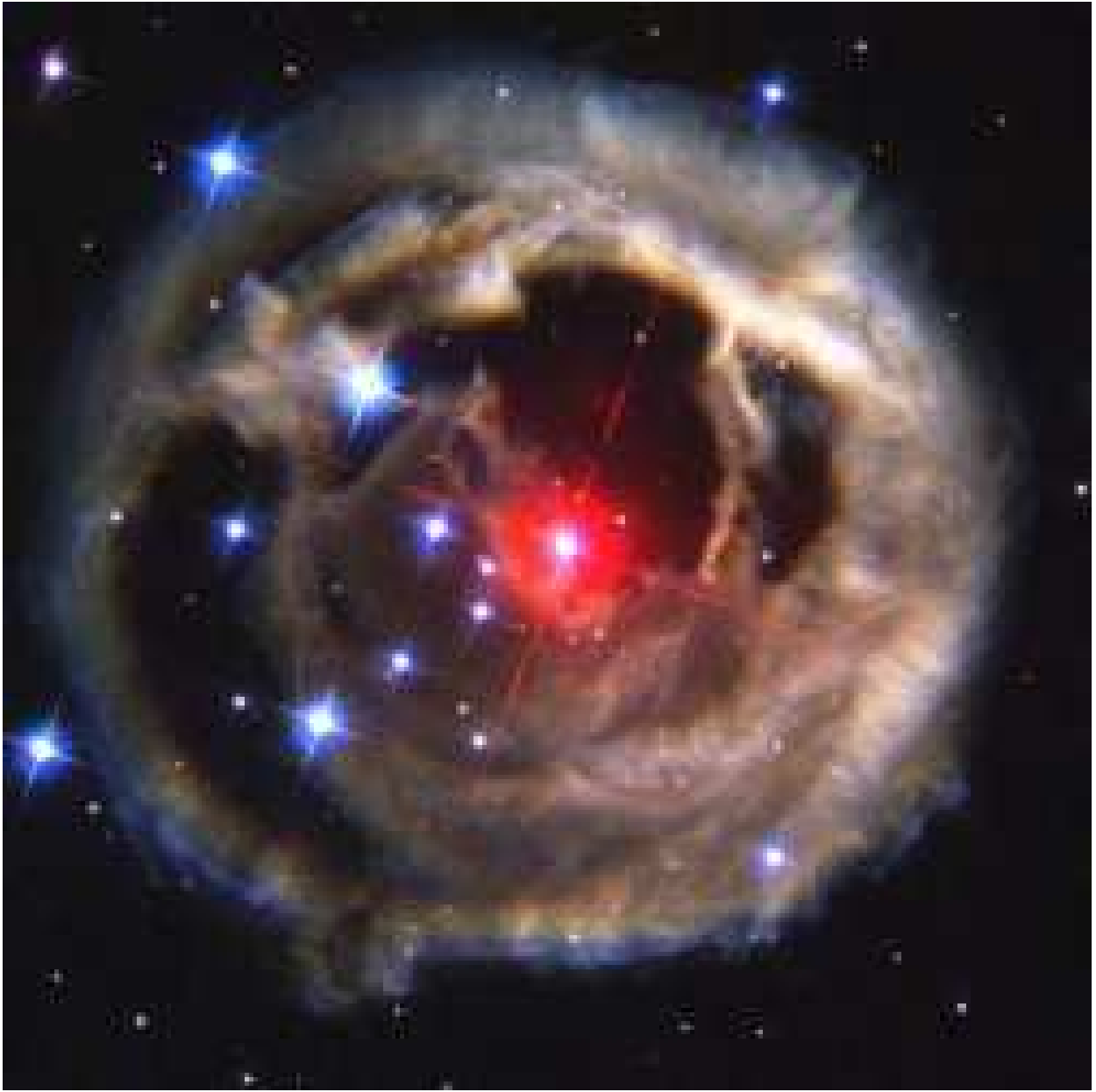}
\end{center}
\figcaption{
Color rendition of the V838~Mon light echo on 2002 December~17, prepared from
the {\it HST\/} $B$, $V$, and $I$ images listed in Tables~1 and 2. Note the
prevalence of blue outer edges and red inner ones, which arise as a consequence
of the echo geometry and the time behavior of the outburst light, as described
in the text. The image is $90''$ high and has north at the top and east on the
left.
}
\end{figure}

\begin{figure}
\begin{center}
\includegraphics[width=4.15in]{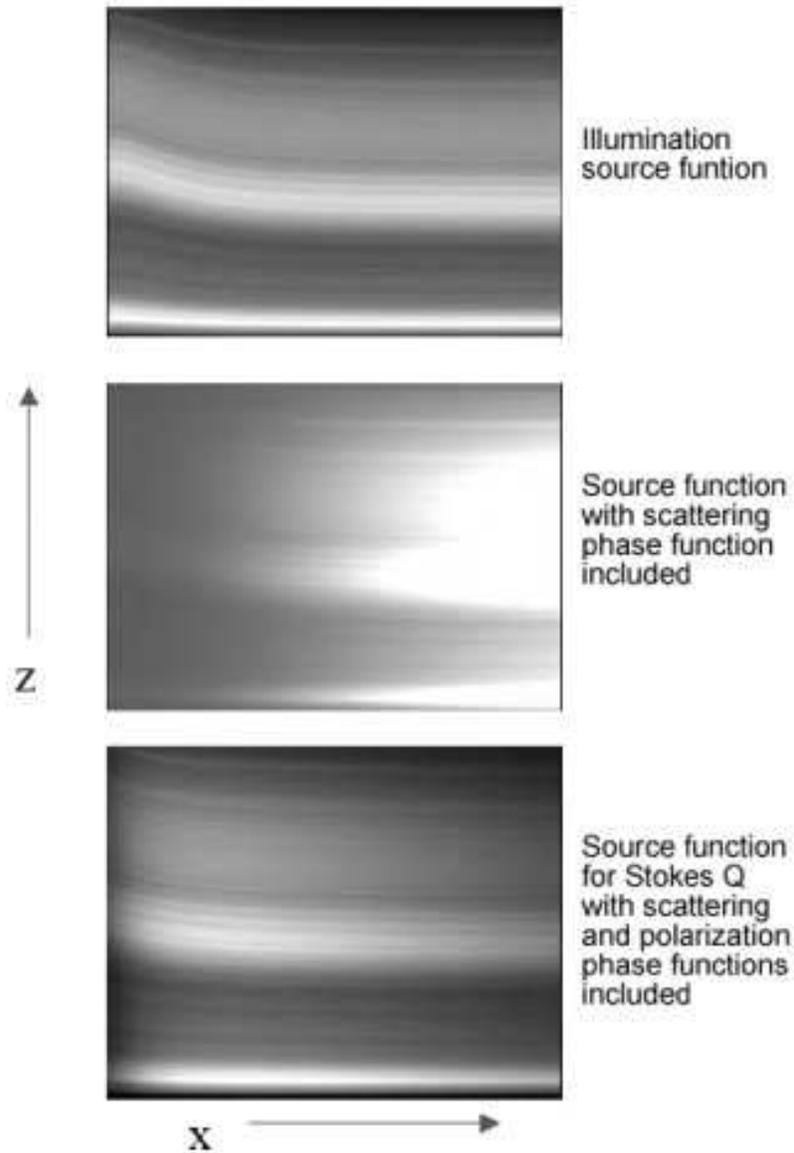}
\end{center}
\figcaption{
Source functions for the V838~Mon light echo, illustrated for a particular
stellar distance (6.1~kpc) and epoch (2002 December 17). In each panel the
source functions are represented by brightness as functions of projected $x$
distance from the star, and $z$ depth in the line of sight on a linear distance
scale from the largest time lag (bottom of each panel) to the smallest lag (top
of each panel). The top panel shows the illumination function from the outburst
light curve; the middle panel also includes the dust scattering function; and
the bottom panel adds the polarization function. See text for details. Such
source functions were constructed for all four epochs and for a large range of
plausible stellar distances.
}
\end{figure}

\begin{figure}
\begin{center}
\includegraphics[width=6in]{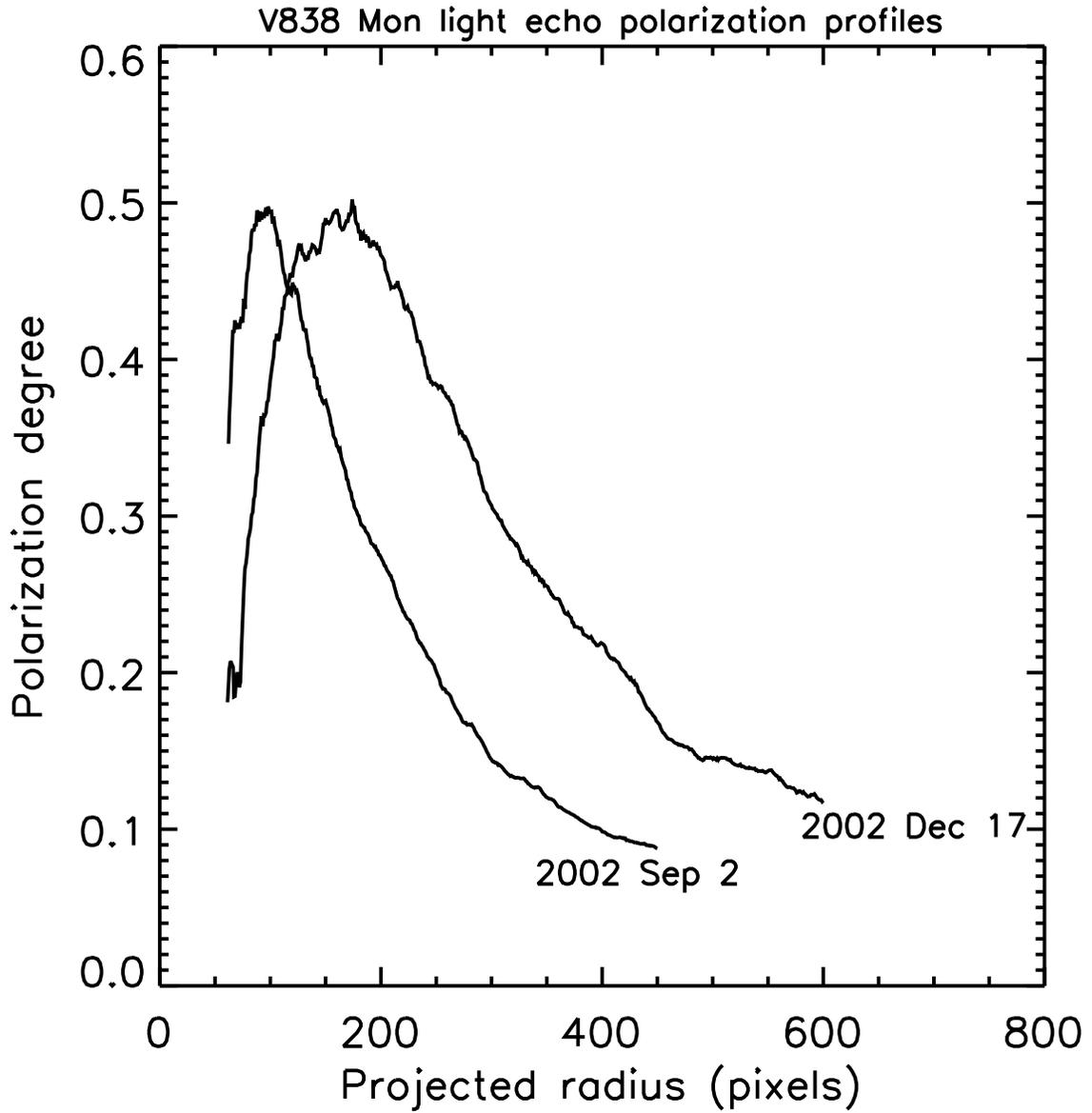}
\end{center}
\figcaption{
Azimuthally averaged polarization profiles for 2002  September and December,
including only data for the regions illustrated in Fig.~4.
}
\end{figure}

\begin{figure}
\begin{center}
\includegraphics[width=6in]{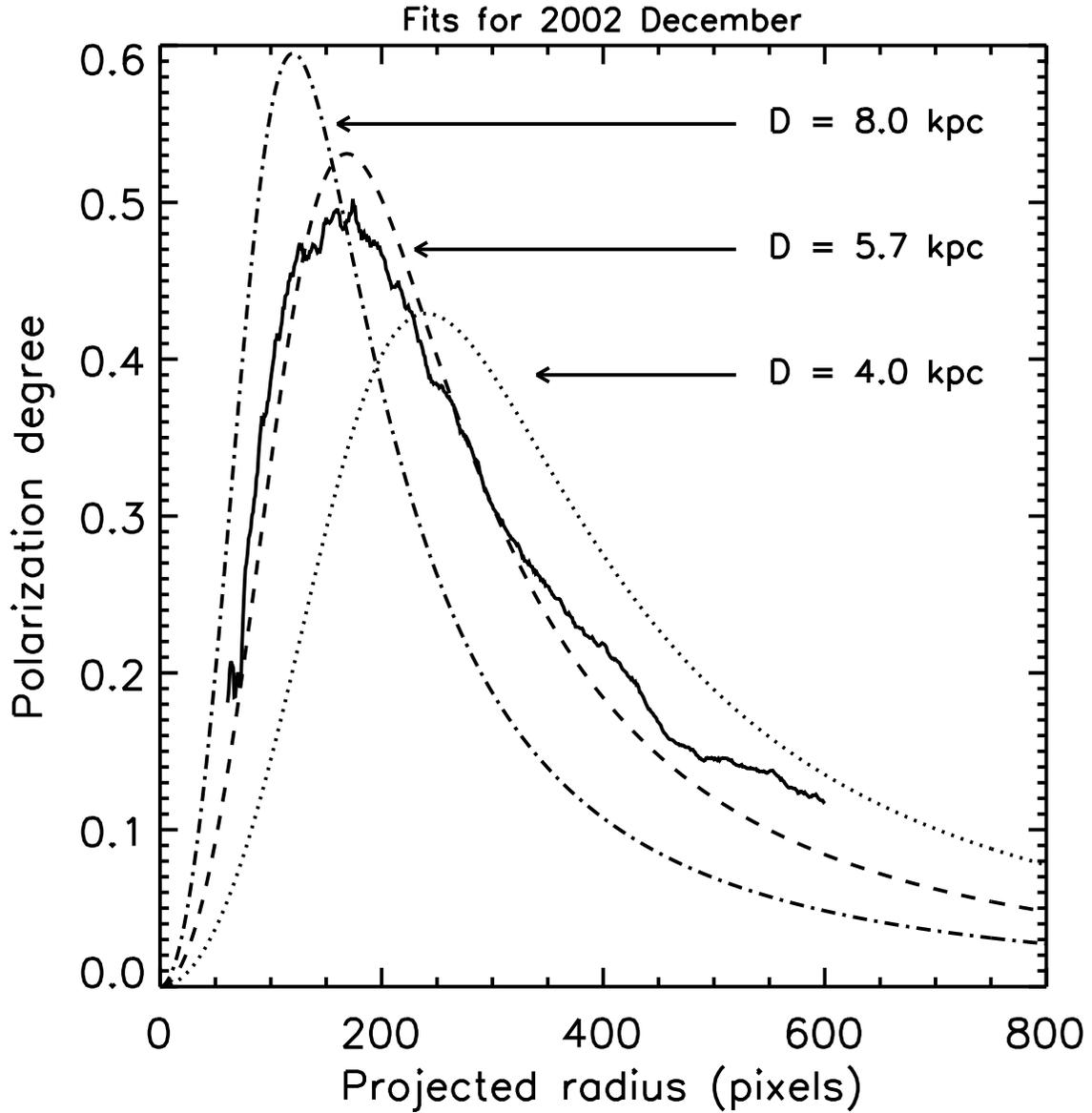}
\end{center}
\figcaption{
Azimuthally averaged polarization data for 2002 December (noisy line) compared
with numerical model Rayleigh-type polarization scattering functions for
distances of 8, 5.7, and 4~kpc (smooth curves). The best fit is for 5.7~kpc.
}
\end{figure}

\begin{figure}
\begin{center}
\includegraphics[width=6in]{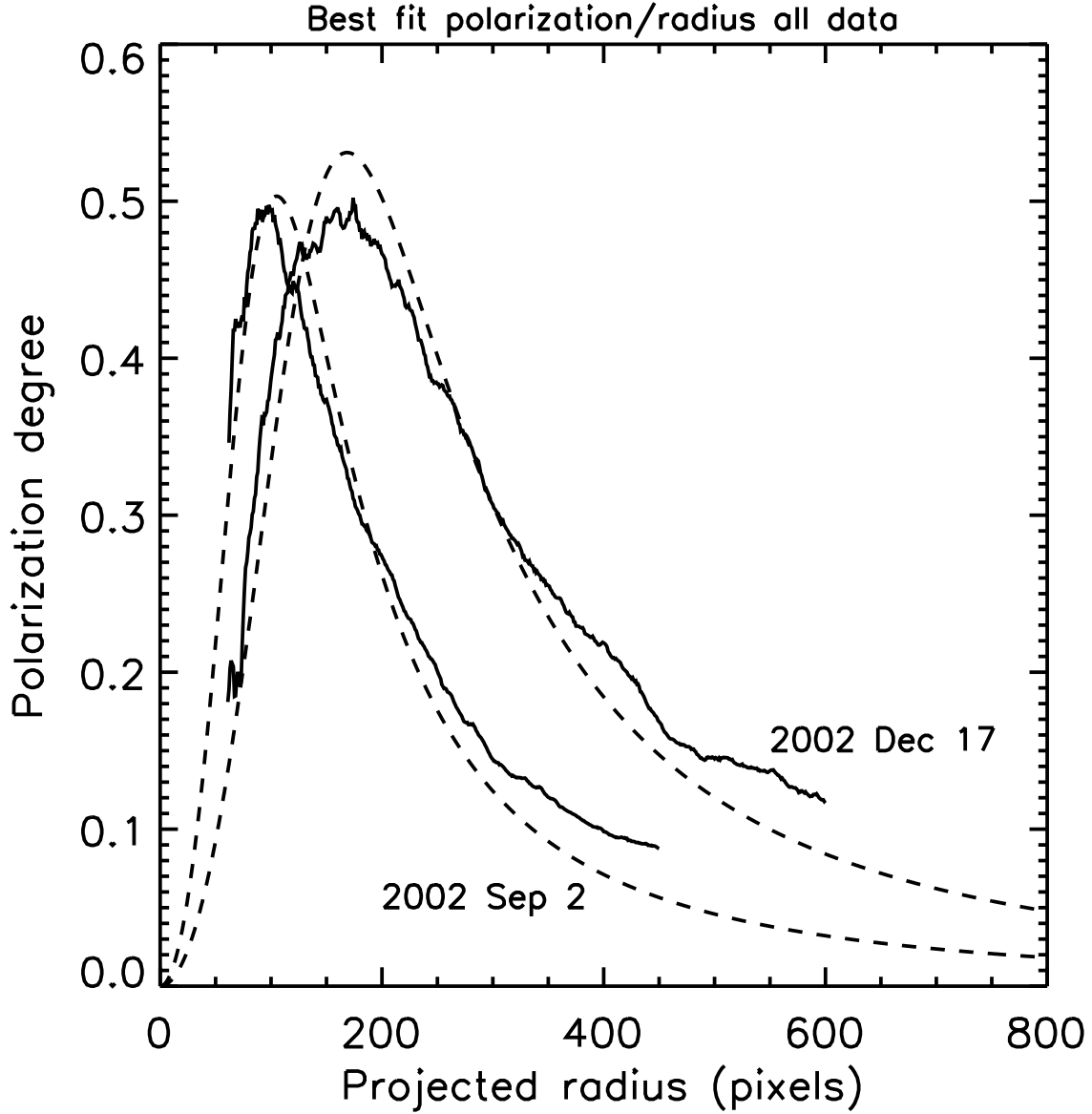}
\end{center}
\figcaption{
Azimuthally averaged polarization data for 2002 December (right-hand noisy line)
with best-fit Rayleigh-type polarization scattering function for a distance of
5.7~kpc (right-hand smooth curve), both repeated from Fig.~10. Also shown are
the data (left-hand noisy line) and a similar fit (left-hand smooth curve) for
the 2002 September polarimetric data, which yield a best-fit distance of
5.6~kpc.
}
\end{figure}

\begin{figure}
\begin{center}
\includegraphics[width=6in]{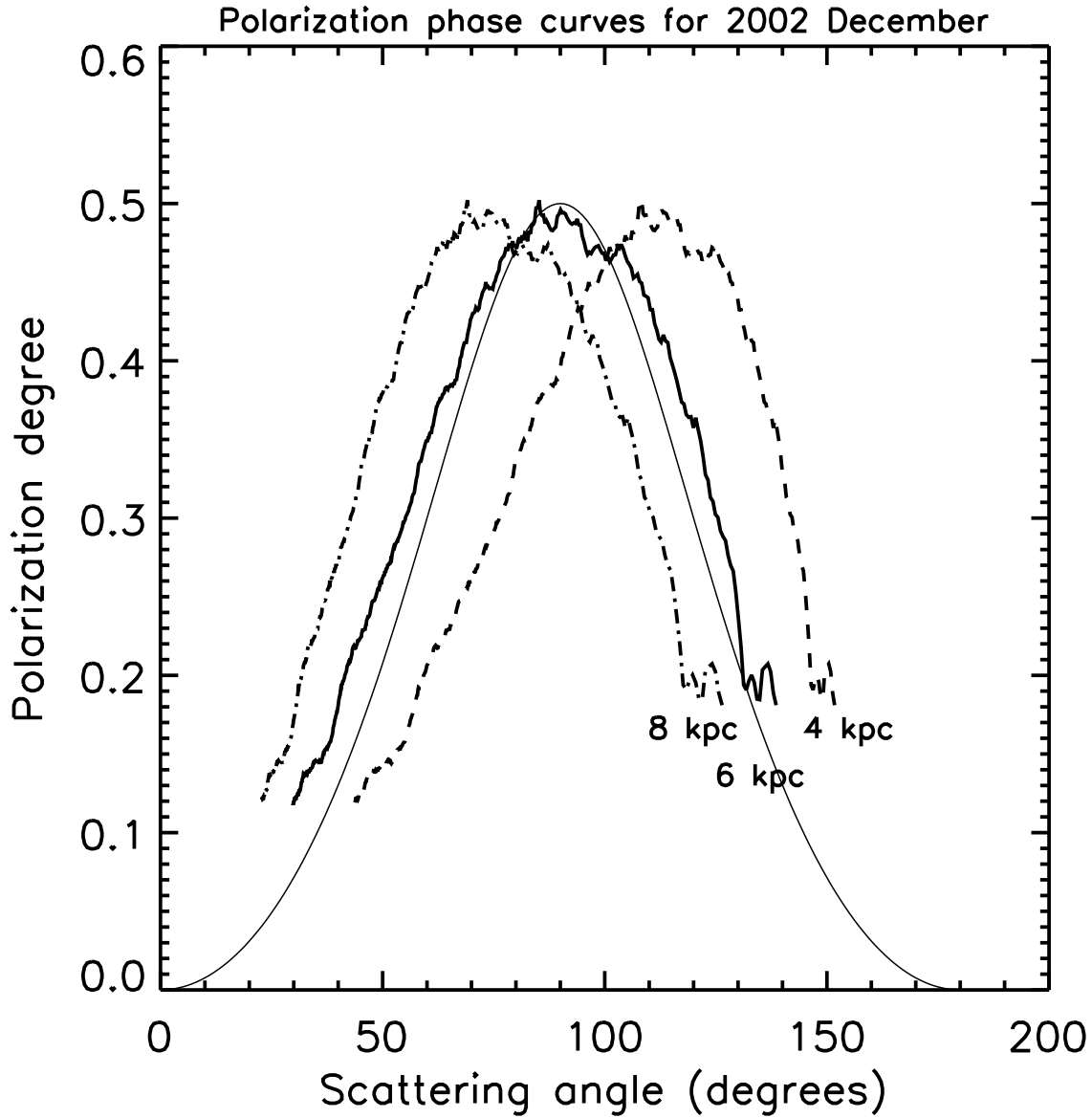}
\end{center}
\figcaption{
Linear polarization degree plotted as a function of scattering angle, as derived
for the 2002 December data and assumed distances of 8~kpc (inner curve), 6~kpc
(middle noisy curve), and 4~kpc (outer curve). If V838~Mon lies at a distance of
4~kpc, peak linear polarization occurs at a scattering angle of
$\sim$$110^\circ$. If it is at a distance of 8~kpc, the peak is at an angle of
$\sim$$75^\circ$. For reference the smooth curve shows a standard Rayleigh
polarization curve given by $p/p_{\rm max} = (1-
\cos^2\theta)/(1+\cos^2\theta)$, with $p_{\rm max}=0.5$, which peaks at a
scattering angle of $90^\circ$.
}
\end{figure}

\begin{figure}
\begin{center}
\includegraphics[width=6in]{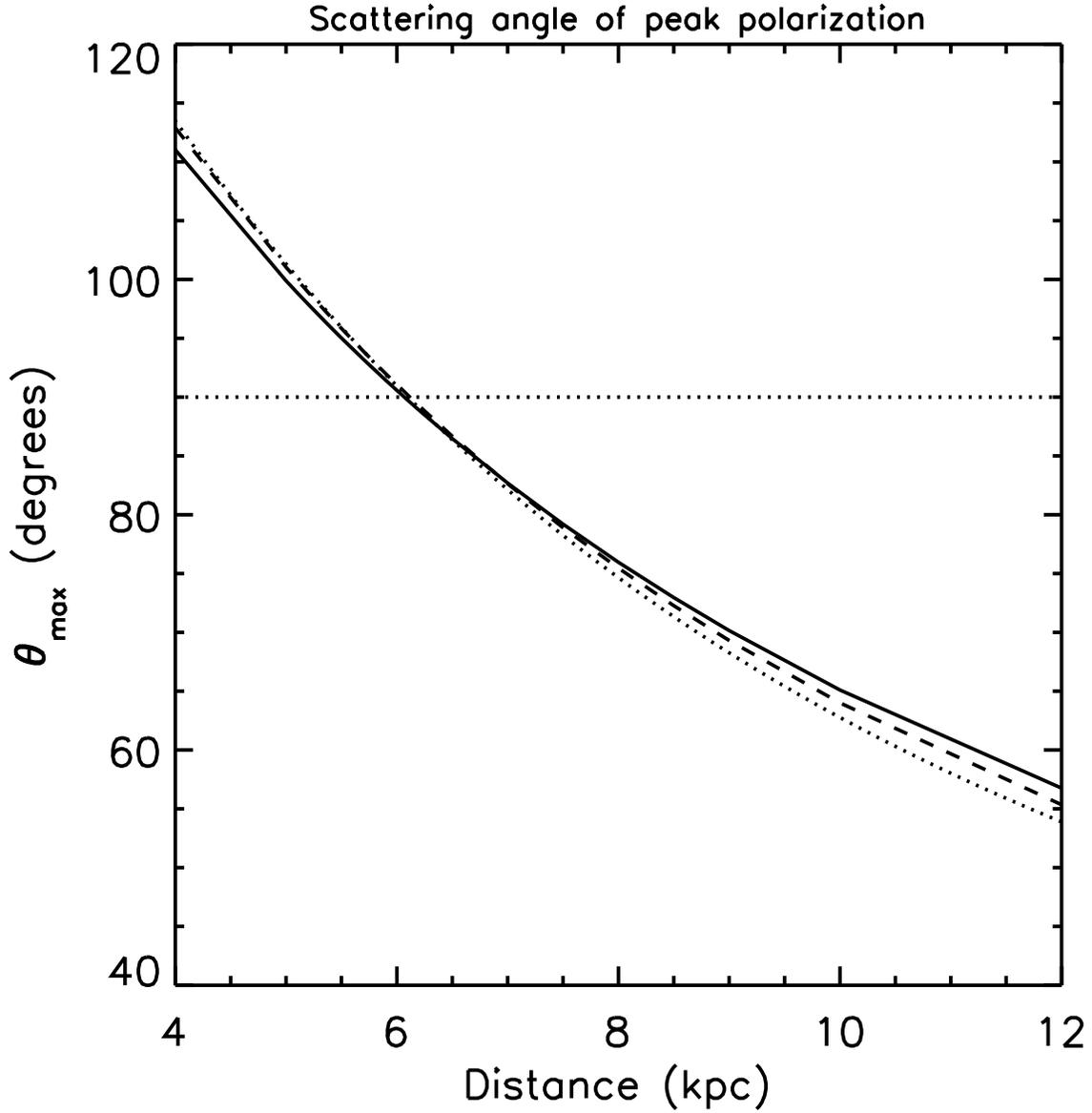}
\end{center}
\figcaption{
Derived scattering angle for peak polarization as a function of distance. The
solid line shows the result for the 2002 December data, and the dashed line for
the 2002 September data. The dotted curve is the analytic solution for an
instantaneous outburst with $D_{90}=6.1$~kpc. The horizontal dotted line shows a
$90^\circ$ scattering angle, which intersects both observed curves at a distance
of 6.1~kpc.
}
\end{figure}

\begin{figure}
\begin{center}
\includegraphics[width=6in]{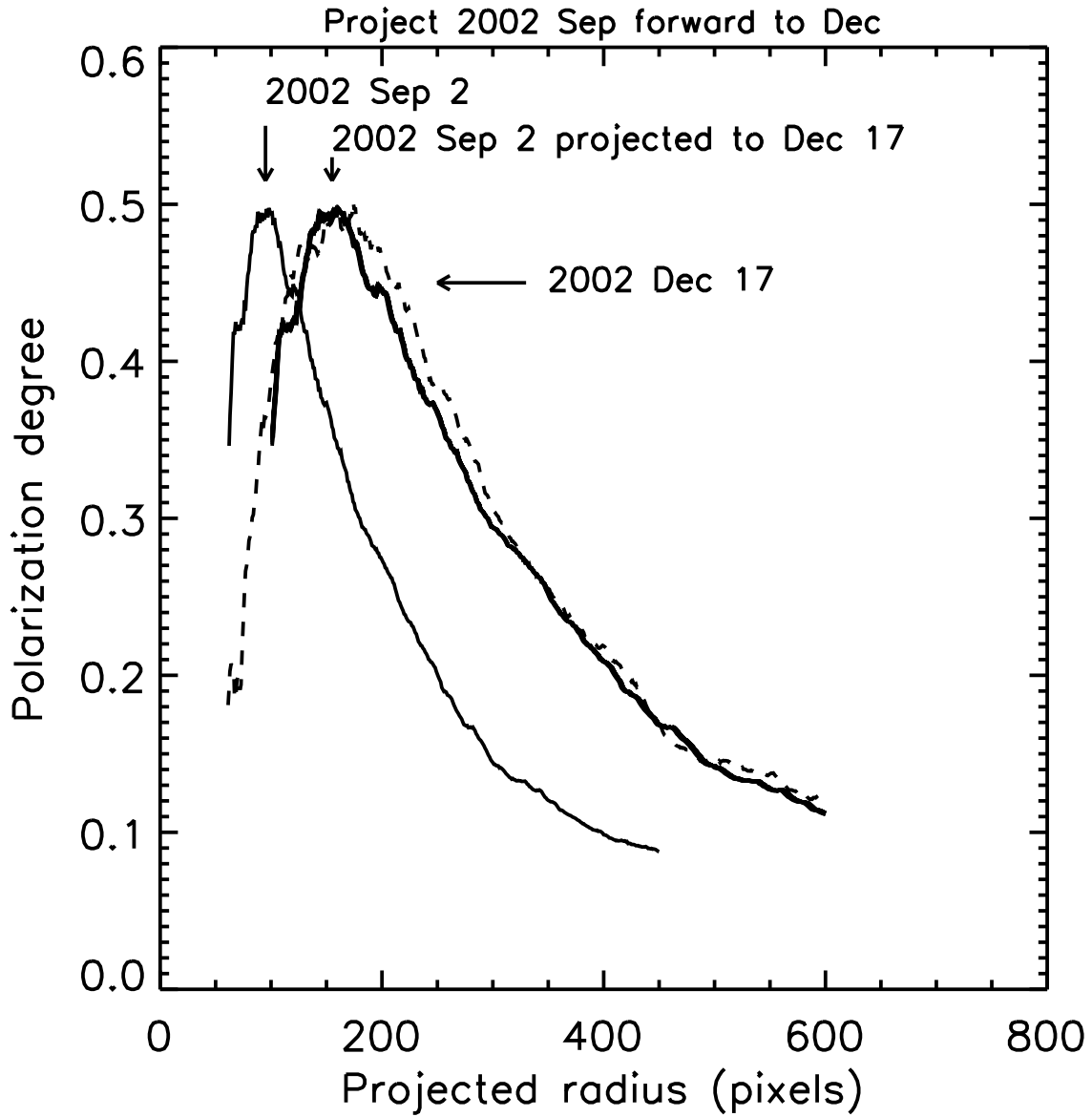}
\end{center}
\figcaption{
Result of mapping the 2002 September polarization profile forward to its
equivalent in 2002 December, using the phase function (i.e., the relationship
between polarization and scattering angle) implied by the September data. The
left-hand curve is the observed polarization profile in 2002 September. The
right-hand curve is the profile predicted for 2002 December using the
September phase function. The dashed line on the right is the actual December
data.
}
\end{figure}

\begin{figure}
\begin{center}
\includegraphics[width=6in]{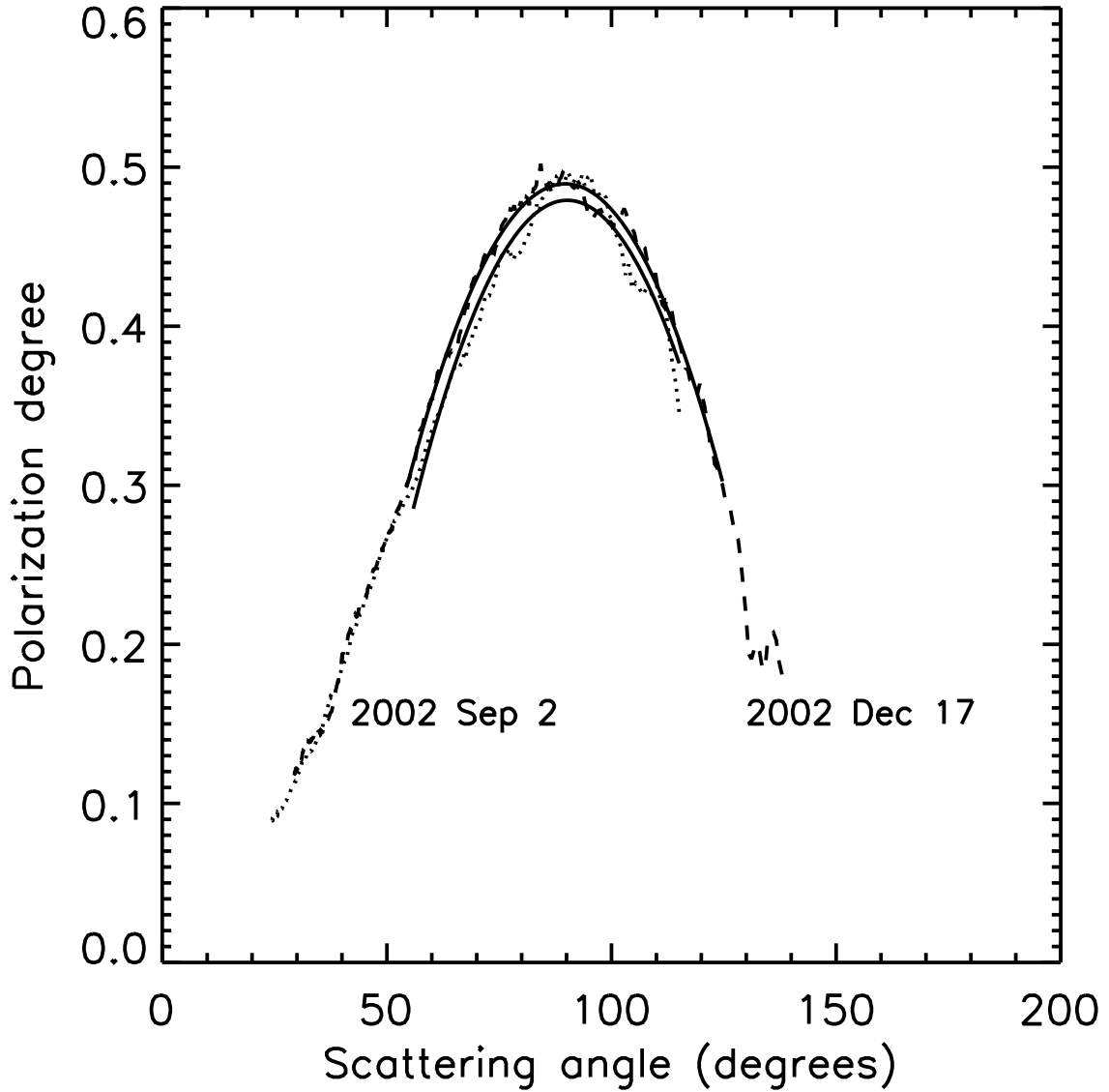}
\end{center}
\figcaption{
Polarization as a function of scattering angle (``polarization phase curves'')
derived from the 2002 September and December data (dotted and dashed lines,
respectively). The adopted distance is 6.1~kpc, which yields a peak polarization
at a scattering angle of $90^\circ$. The solid curves show parabolic
least-squares fits to the two data sets for values of $p>0.3$ only.
}
\end{figure}

\begin{figure}
\begin{center}
\includegraphics[width=6in]{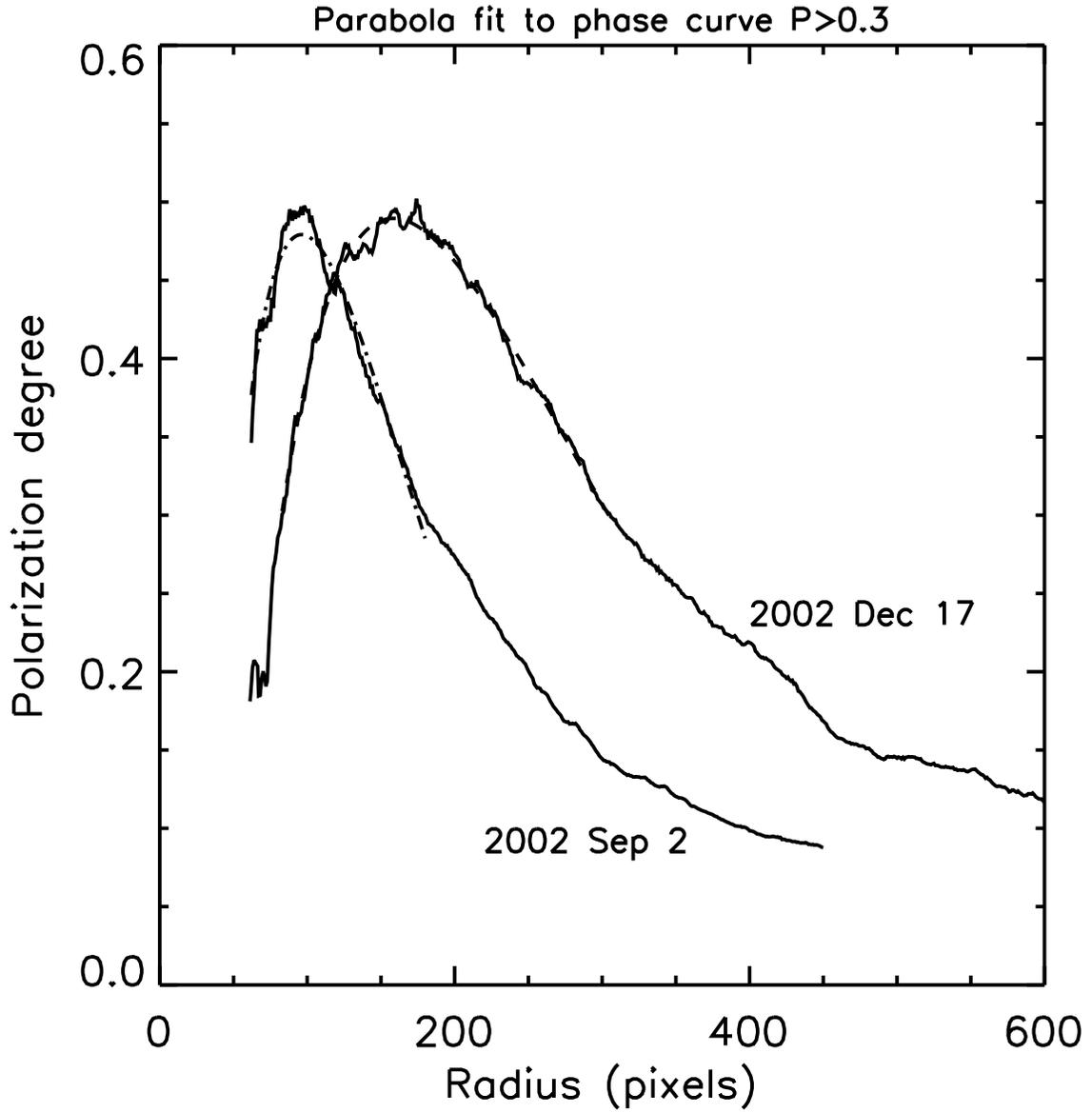}
\end{center}
\figcaption{
Polarization as a function of radius for the 2002 September and December epochs,
calculated for the phase functions represented by the parabolic fits shown in
Fig.~15 (smooth curves). The actual data are shown as noisy curves.
}
\end{figure}

\begin{figure}
\begin{center}
\includegraphics[width=5.25in]{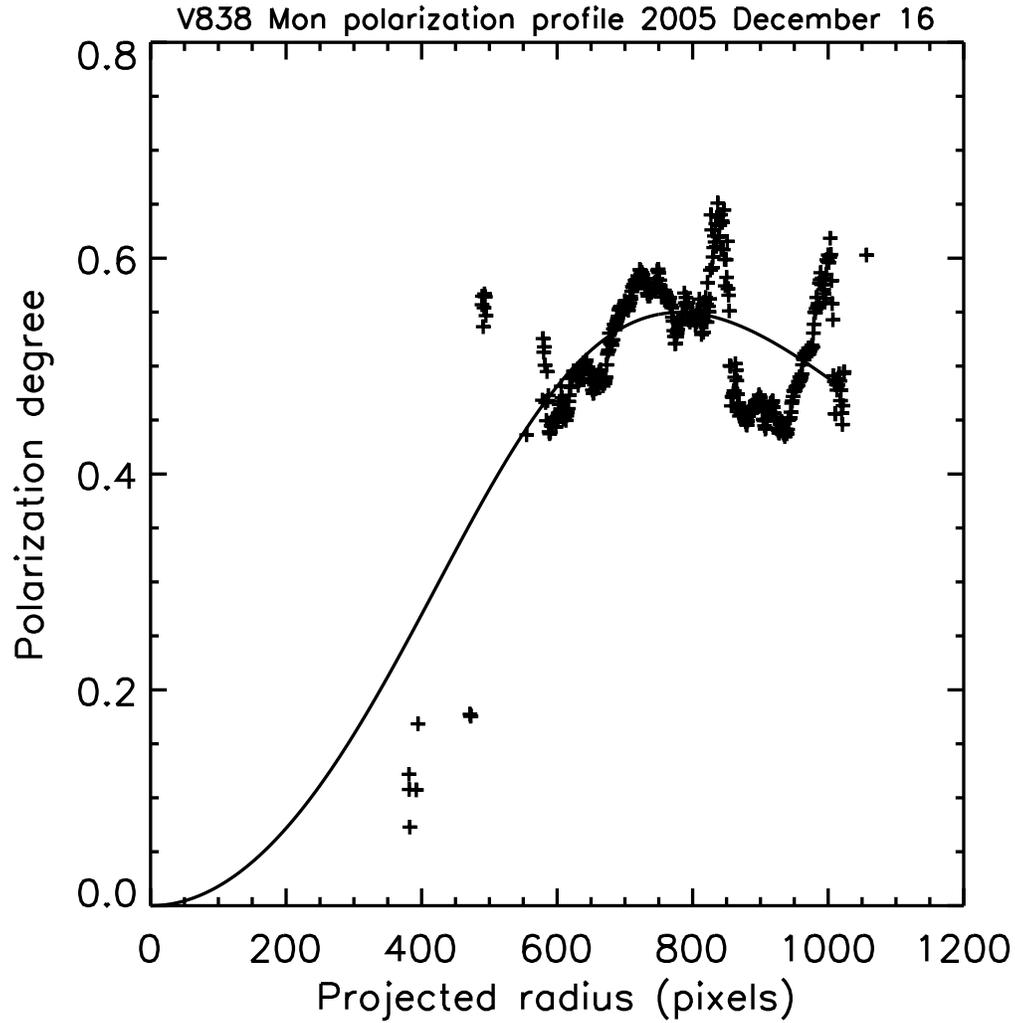}
\end{center}
\figcaption{
The crosses show the azimuthally averaged polarization profile for 2005
December 16, based on the masked data shown in Fig.~4. The smooth curve is the
predicted profile, calculated using the polarization phase function of 2002
December 17 and an assumed distance of 6.1~kpc. Here $p_{\rm max}$ was set to
0.55 rather than 0.5, but otherwise the observations are reasonably consistent
with expectation.
}
\end{figure}

\clearpage

\begin{deluxetable}{llcll}
\tablewidth{0 pt}
\tablecaption{Log of \HST\/ ACS WFC Polarimetric Observations of V838~Mon}
\tablehead{
\colhead{Dataset} &
\colhead{UT Date} &
\colhead{Exposure (s)} &
\colhead{Filters} &
\colhead{Proposal ID \& PI} }
\startdata
J8GG01011 & 2002 Apr 30 & 506 & F435W;POL0UV   & \phn9587 Starrfield \\
J8GG01021 & 2002 Apr 30 & 506 & F435W;POL60UV  & \phn9587 Starrfield  \\
J8GG01031 & 2002 Apr 30 & 506 & F435W;POL120UV & \phn9587 Starrfield \\
\noalign{\smallskip}
J8GG02011 & 2002 May 20 & 506 & F435W;POL0UV   & \phn9587 Starrfield \\
J8GG02021 & 2002 May 20 & 506 & F435W;POL60UV  & \phn9587 Starrfield \\
J8GG02031 & 2002 May 21 & 506 & F435W;POL120UV & \phn9587 Starrfield \\
\noalign{\smallskip}
J8GL03011 & 2002 May 20 & 362 & F606W;POL0V    & \phn9588 Bond\\
J8GL03021 & 2002 May 20 & 362 & F606W;POL60V   & \phn9588 Bond \\
J8GL03031 & 2002 May 20 & 362 & F606W;POL120V  & \phn9588 Bond \\
\noalign{\smallskip}
J8JY04011 & 2002 Sep  2 &  960 & F606W;POL0V & \phn9694 Bond \\
J8JY04021 & 2002 Sep  2 &  960 & F606W;POL60V & \phn9694 Bond \\
J8JY04031 & 2002 Sep  2 &  960 & F606W;POL120V & \phn9694 Bond \\
\noalign{\smallskip}
J8JY06011 & 2002 Dec 17 &  910 & F606W;POL0V & \phn9694 Bond \\
J8JY06021 & 2002 Dec 17 &  910 & F606W;POL60V & \phn9694 Bond \\
J8JY06031 & 2002 Dec 17 &  910 & F606W;POL120V & \phn9694 Bond \\
\noalign{\smallskip}
J9BS06030 & 2005 Dec 16 &  \llap{1}374 & F606W;POL0V & 10618 Bond \\
J9BS06040 & 2005 Dec 16 &  \llap{1}374 & F606W;POL60V & 10618 Bond \\
J9BS06050 & 2005 Dec 16 &  \llap{1}374 & F606W;POL120V & 10618 Bond \\
\enddata
\end{deluxetable}

\begin{deluxetable}{llclc}
\tablewidth{0 pt}
\tablecaption{Log of \HST\/ ACS WFC Non-Polarimetric Observations Used in This
Paper}
\tablehead{
\colhead{Dataset} &
\colhead{UT Date} &
\colhead{Exposure (s)} &
\colhead{Filter} &
\colhead{Proposal ID \& PI} }
\startdata
J8JY06041 & 2002 Dec 17 &  1070 & F435W & 9694 Bond \\
J8JY06051 & 2002 Dec 17 &  1540 & F435W & 9694 Bond \\
J8JY06NOQ & 2002 Dec 17 & \phn100 & F814W & 9694 Bond \\
J8JY06OQQ & 2002 Dec 17 & \phn100 & F814W & 9694 Bond \\
\noalign{\smallskip}
J9BS06020 & 2005 Dec 16 & 3262 & F606W & \llap{1}0618 Bond \\
\enddata
\end{deluxetable}

\begin{deluxetable}{lcc}
\tablewidth{0 pt}
\tablecaption{Position Angles and Throughputs for ACS WFC Polarizers}
\tablehead{
\colhead{Polarizer} &
\colhead{Position Angle\tablenotemark{a}} &
\colhead{Throughput\tablenotemark{b}} 
}
\startdata
POL0V   & \llap{$-$}$38\fdg2$ & 1.000 \\
POL60V  & $21\fdg8$  & 0.979 \\
POL120V & $81\fdg8$  & 1.014 \\
\enddata
\tablecomments{Values are adopted from Biretta et al.\ 2004.}
\tablenotetext{a} {Position angle of the electric vector measured
counter-clockwise from the spacecraft V3 axis.}
\tablenotetext{b} {Relative throughputs for polarizers combined
with the F606W bandpass filter.}
\end{deluxetable}

\end{document}